\documentclass[11pt, notitlepage]{article} 
\usepackage{amsmath, amsfonts, amssymb, graphicx, cases, url, bbm, multicol, multirow, amsthm}
\usepackage[letterpaper]{geometry}
\usepackage{caption}
\usepackage{subfigure}
\usepackage{authblk}
\usepackage{epstopdf}
\usepackage[noadjust]{cite}
\DeclareMathOperator*{\argmax}{arg\,max}

\newcommand\blfootnote[1]{%
  \begingroup
  \renewcommand\thefootnote{}\footnote{#1}%
  \addtocounter{footnote}{-1}%
  \endgroup
}

\usepackage[ruled]{algorithm2e}

\theoremstyle{definition}
\newtheorem{exmp}{Example}[section]

\SetAlFnt{\small}
\SetAlCapFnt{\small}
\SetAlCapNameFnt{\small}
\SetAlCapHSkip{0pt}
\IncMargin{-\parindent}

\ifodd 0
\newcommand{\rev}[1]{{\color{blue}#1}} 
\newcommand{\com}[1]{\textbf{\color{red}(COMMENT: #1)}} 
\newcommand{\clar}[1]{\textbf{\color{green}(NEED CLARIFICATION: #1)}}

\else
\newcommand{\rev}[1]{#1}
\newcommand{\com}[1]{}
\newcommand{\clar}[1]{}
\fi

\begin{document}


\title{Efficient Sensor Fault Detection Using Group Testing}

\author[1]{CHUN LO}
\author[1]{YECHAO BAI}
\author[1]{MINGYAN LIU}
\author[2]{JEROME P. LYNCH}
\affil[1]{Electrical Engineering and Computer Science, University of Michigan, Ann Arbor}
\affil[2]{Civil and Environmental Engineering, University of Michigan, Ann Arbor}

\renewcommand\Authands{ and }

\maketitle

\blfootnote{A preliminary version of part of this work appeared in the IEEE International Conference on Computing in Sensor Systems (DCOSS) 2013 \cite{lo2013efficient}.
\newline CHUN LO: chunlo@umich.edu, YECHAO BAI: ychbai@nju.edu.cn, MINGYAN LIU: mingyan@umich.edu, JEROME P. LYNCH: jerlynch@umich.edu}

\begin{abstract}
When faulty sensors are rare in a network, diagnosing sensors individually is inefficient. This study introduces a novel use of concepts from group testing and Kalman filtering in detecting these rare faulty sensors with significantly fewer number of tests. By assigning sensors to groups and performing Kalman filter-based fault detection over these groups, we obtain binary detection outcomes, which can then be used to recover the fault state of all sensors. We first present this method using combinatorial group testing. We then present a novel adaptive group testing method based on Bayesian inference. This adaptive method further reduces the number of required tests and is suitable for noisy group test systems. Compared to non-group testing methods, our algorithm achieves similar detection accuracy with fewer tests and thus lower computational complexity. Compared to other adaptive group testing methods, the proposed  method achieves higher accuracy when test results are noisy. We perform extensive numerical analysis using a set of real vibration data collected from the New Carquinez Bridge in California using an 18-sensor network mounted on the bridge. We also discuss how the features of the Kalman filter-based group test can be exploited in forming groups and further improving the detection accuracy.
\end{abstract}

\section{Introduction}

Wireless sensor networks (WSNs) have been successfully used in many applications such as structural health monitoring \cite{lynch2007overview}, environmental monitoring \cite{WSN_agriculture} and vehicle tracking \cite{cartel}.
With the increasing use of small, low power and low cost sensors, it has also become increasingly critical to ensure the accuracy and integrity of the measured data as low cost sensors are error prone while the environment in which they are deployed may be harsh.
Timely detection of malfunctioning sensors in a system allows the operator to correct affected sensor readings and arrange for replacement, both of which can prevent further deterioration of the network, and thus should be an essential functionality of a WSN.

Over the past decade, detection of malfunctioning sensors has been studied extensively in many different application contexts.  Malfunctioning can be classified into two levels. The first is sensor failure, whereby sensors become irresponsive or cease to provide data, see e.g., \cite{Ramanathan:2005:failure:sensor,Staddon:2002:failednode:tracing,Ruiz:2004:Event:Driven}.
The second is sensor faulting, whereby sensors continue to report measurements but the data are intermittently or permanently corrupted.
Sensor fault detection is generally more difficult than sensor failure detection because it is typically harder to assess the accuracy of data than it is to determine its absence.  In this paper our focus is on the former.

Sensor fault detection methods can be further classified as model-based and model-free.
Model-based fault detection methods rely on a model capturing the dynamics of the system being monitored; this model can be obtained either from physical properties of the system (e.g., a state-space model) or from learning the parameters of a designated model (e.g., a Markov or autoregressive model). Both Kobayashi et al. \cite{kobayashi2003application} and Da et al. \cite{dalin:kalman} proposed centralized detection algorithms which assume a state-space model of the system is available, and a bank of Kalman filters is used to detect faulty sensors. Both methods assume there is {\em at most} one faulty sensor at any given time and make use of the remaining sensors as  references. A more detailed and quantitative comparison is given in Section \ref{sec:CGT_result}.
Li et al. \cite{Li:iomodel} proposed an algorithm that requires fault-free sensors be designated \emph{a priori} as reference sensors, and the number of reference sensors is required to be more than the number of uncertain sensors (i.e., those in unknown fault state). Their algorithm constructs analytical relationship between the output of each uncertain sensor and that of all reference sensors, which is then used for detection.
Ricquebourg et al. \cite{Ricquebourg:TBM} 
modeled sensor dynamics using a Markov chain under a transferable belief framework when the whole system is healthy. Once the model is established, any sensor outputs inconsistent with the model are further analyzed using predefined decision rules.
Lo et al. \cite{Lo:2011:reference_free_FD} proposed a decentralized algorithm which is able to identify spike faults in addition to detecting general faults. Under this method pairs of sensors cross-validate each other using their measurements and an autoregressive with exogenous input model (ARX) trained \emph{a priori}; this method does not require reference sensors or \emph{a priori} knowledge of the system model.


Model-free fault detection methods do not require a dynamical model and usually rely on the assumption that sensors in close proximity observe similar dynamics. As a result, the density of the sensors needs to be high relative to the fluctuation of the signals being monitored. 
For instance, Ding et al. \cite{Ding:Localized_F_Tolerant} and Chen et al. \cite{Chen:distributed} suggested similar model-free sensor fault detection methods, where each sensor's output is compared with its neighbors'. 
A sensor that deviates significantly from its neighbors is identified as faulty. 
Koushanfar et al. \cite{koushanfar2003line} proposed a cross-validation based fault detection algorithm that focuses on the impact of a particular sensor's measurement on the consistency of the entire network's measurement, under the assumption that an incorrect measurement will degrade the consistency. This algorithm removes one sensor at a time and evaluates how much the consistency of the system improves. The sensor whose removal improves the system most significantly is regarded as faulty and eliminated and the process is repeated until the system consistency cannot be improved anymore.

All of the above mentioned fault detection methods require the number of tests at least on the order of the size of the network, i.e., $\mathcal{O}(N)$ tests are required, where $N$ is the number of sensors in the network.  Some methods even need $\mathcal{O}(mN)$ (where $m$ is the number of neighbors of a sensor) or $\mathcal{O}(N^2)$ tests. A summary of the detection complexity is given in Table \ref{table:complexity}. For applications using an extremely large number of sensors \cite{example_high_densityWSN}, running a fault detection algorithm can involve a large amount of resources and cause significant delay.
\begin{table}[tb]
\centering
\begin{tabular}{|l|l|l|}
\hline
Method type  & Complexity & Condition needed \\
\hline
\textbf{Model-based}: &&\\
Kobayashi et al.\cite{kobayashi2003application}    & $\mathcal{O}(N)$ & At most one faulty sensor \\
Da et al.\cite{dalin:kalman}            & $\mathcal{O}(N)$ & At most one faulty sensor \\
Lo et al.\cite{Lo:2011:reference_free_FD} & $\mathcal{O}(N)$ &  \\
Li et al.\cite{Li:iomodel}          & $\mathcal{O}(N)$ &  Reference sensor \\
Ricquebourg et al.\cite{Ricquebourg:TBM}   & $\mathcal{O}(N)$ & \\
\hline
\textbf{Model-free}:  &&\\
Ding et al.\cite{Ding:Localized_F_Tolerant}         & $\mathcal{O}(mN)$     & $m=$ \# of neighbors \\
Chen et al.\cite{Chen:distributed}         & $\mathcal{O}(mN)$     & $m=$ \# of neighbors \\
Koushanfar et al.\cite{koushanfar2003line}   & $\mathcal{O}(N^2)$      &  \\
Blough et al.\cite{blough1989fault}       & $\mathcal{O}(N\log N)$      &  \\
\hline
\end{tabular}
\caption{Summary of existing methods}
\label{table:complexity}
\end{table}

We observe that while certain regional effects or catastrophic failure may result in a large number of faulty sensors at the same time, in the absence of such systemic problems and during normal operation faults occur randomly and sporadically. 
This motivates us to seek lower complexity fault detection methods when faults may be rare and sparse.

Toward this end, we introduce a novel use of group testing techniques combined with Kalman filtering in detecting faulty sensors in a network.  Assuming that the underlying system being monitored may be represented in a linear dynamical system framework and that sensor faults are relatively rare, our goal is to reduce the number of required tests given requirements on detection and false positive probabilities. There have been a few studies on using group testing to detect malfunctioning sensors; they generally differ in the testing/detection methods.
For instance, Goodrich and Hirschberg \cite{goodrich2006efficient} 
evaluates a group of sensors by counting the number of responses from the group to a broadcast query (thus only applicable to sensor failure detection rather than fault detection), while To\v{s}i\'{c} \emph{et al.} \cite{tosic2012distributedGT} 
uses an unspecified dissimilarity comparison of neighboring sensors' measurements. \rev{Our work differs from the former in that we focus on detecting faulty sensors which are still responsive to queries, and differs from the latter in that we do not assume that sensors are highly correlated or that neighboring sensors have similar measurements.}

Our approach consists of the following two components: the selection of a test group (also referred to as a test pool), and a Kalman filtering based testing/detection procedure over this group of sensors, which determines whether there exists at least one faulty sensor in this group.  These two steps are repeated till desired performance criteria have been achieved.  There are in general two ways of selecting the test groups.  The first is open-loop, whereby the entire set of test groups are selected prior to performing any tests (this is done randomly in our study); this will be referred to as the combinatorial group testing (CGT) method.  The second is closed-loop, whereby each test group is selected adaptively based on outcomes of previous tests (this adaptive section is done using the standard criteria of uncertainty reduction maximization in our study); this will be referred to as the Bayesian group testing (BGT) method.  Both methods will be examined in this study.
\rev{We will further consider the detection performance of Kalman filtering, and use such understanding in determining the selection of test groups under the Bayesian group testing method; this will be referred to as the Kalman filtering-enhanced Bayesian group testing method (KF-BGT).} It should be emphasized that under all these methods the group tests (the second component) themselves are performed via Kalman filtering; they simply differ in how the test groups are selected (the first component).

Existing adaptive group testing methods generally assumes error-free detection, thus an entire group of sensors is removed from further consideration when the test result is negative.  Examples include Hwang's generalized binary splitting algorithm \cite{hwang1972binary}, Allemann's split-and-overlap algorithm \cite{allemann2003BGT_split} and Du \emph{et al.}'s competitive GT algorithm \cite{du1993combinatorial}.  Test errors have been considered in the literature of compressive sensing, (e.g., see \cite{malloy2012near_O_CS,ji2008bayesianCS}), which is closely related to group testing.  However, these adaptive methods are not directly applicable to group testing as the latter is given by a Boolean operation whereas compressive sensing based test results are given by a linear operation.  Our study further differs from both because our test results are given by a Kalman filtering based detection procedure (neither Boolean nor a linear operation), which is noisy and its result dependent on the design of the test and the detector.  This raises significant challenge that we will address in this paper.

The remainder of the paper is organized as follows: Section 2 reviews the main concepts used in the group testing-based fault detection algorithm. The detailed methodology of the detection algorithm based on CGT is explained in Section 3. The Bayesian group testing method BGT is described in Section 4. Section 5 describes the experimental set up and the nature of a set of bridge vibration data we use for numerical evaluation. The performance of the CGT and BGT methods on the bridge vibration data is presented in Sections 6 and 7, respectively. The analysis of the Kalman filtering-enhanced version KF-BGT methods is presented in Section 8. Section 9 concludes the paper.

\section{Preliminaries} \label{sec:preliminaries}
In this section we review two main concepts used in our fault detection algorithm. The first is group testing, the goal of which is to identify sparse faulty items with fewer number of tests than the total number of items. The second concept is Kalman filtering, which is able to produce optimal state estimation for a linear dynamical system.

\subsection{Group Testing}
Consider a large number of items of which a few are defective. If each item is tested individually, the cost can be high (linear in the total number of items).  However, if it is possible to determine the existence of a defective item in a group via a single {\em group test}, then performing a sequence of group tests over different subsets of these items can potentially lead to much fewer number of tests and thus much lower cost.  This is the main idea of group testing; it was first proposed by Dorfman \cite{dorfman1943detection} during World War II for detecting syphilis amongst soldiers.

Consider a length $N$ signal $S$ which is $d$ sparse: this means $S$ has at most $d$ non-zero entries that correspond to the defective items and $d\ll N$. As the ``true" signal dimension (i.e., $d$) is smaller than $N$, it is conceivable that signal $S$ can be acquired with $M<N$ measurements. In group testing paradigm, signal $S$ is measured $M$ times in the form of $W=\Phi S$, where $\Phi$ is the measurement matrix of size $M \times N$. The arithmetic is boolean, meaning that the multiplication is  logical AND and addition logical OR.  If these operations are noisy, then the group test results are given by $Z$ rather than $W$, with $P(Z_i=1|W_i=0)=\alpha$ and $P(Z_i=0|W_i=1)=\beta, \ \forall i$, denoting the two types of errors. The goal of group testing is to design $\Phi$ such that $S$ can be reconstructed from $Z$ (i.e., we can find the $d$ defective items) with sufficiently low error probabilities.


We now describe this in the context of a network of $N$ sensors, of which at most $d$ are faulty. Let vector $S$ represent the true {\em fault state} of the sensors in the network, where  $S_i=0$ if sensor $i$ is normal and $S_i=1$ if sensor $i$ is faulty.  The $i^{th}$ row of the 0-1 matrix $\Phi$  represents the set of sensors involved in the $i^{th}$ test, and is called a test group/pool denoted by $\Phi_i$; the number of rows equals the number of tests.  Finally, the vector $Z$ represents the result of the group tests.  Below is a toy example of $\Phi S = Z$: 
\begin{exmp}\label{exam:ex1}
\begin{equation*}
    \begin{bmatrix}
        0 & 1 & 0 & 0 & 1 & 1 \\
        0 & 0 & 1 & 1 & 0 & 1 \\
        1 & 0 & 0 & 1 & 1 & 0
    \end{bmatrix}
    \begin{bmatrix}
        0\\
        1\\
        0\\
        0\\
        0\\
        0\\
    \end{bmatrix}
    =
    \begin{bmatrix}
        1\\
        0\\
        1\\
    \end{bmatrix}
\end{equation*}
\end{exmp}
In this example, there are $6$ sensors; sensor $2$ is faulty.  A total of $3$ group tests are performed: sensors \{2, 5, 6\} are included in the first test (first row of $\Phi$), and so on. The test result shows correctly that the first group contains at least one faulty sensor and the second group has none, but declares incorrectly that the third group contains a faulty sensor. In a fault detection setting, $S$ is unknown while $\Phi$ is known by design and $Z$ is known by observing the test results.  $\Phi$ and $Z$ are then used to reconstruct $S$.
As mentioned in the introduction, group-testing a set of sensors in our context is far more complicated than a simple boolean operator, noise-free or noisy.  
To use this group testing framework in practice, we must specify what a ``group test'' entails, and how to actually obtain values in the $Z$ vector.  This is addressed by a novel use of Kalman filtering detailed next.

\subsection{Kalman Filter Based Group Test}
\label{sec:KF}
The Kalman filter \cite{maybeck1979stochastic} is an algorithm which takes a series of noisy inputs and iteratively calculates a statistically optimal estimate of the state of an underlying linear dynamical system.
More specifically, consider a linear dynamical system given by the following state-space model \cite{maybeck1979stochastic}:
\begin{align}
    \textbf{X}_{k+1} &=\textbf{AX}_k+\textbf{BU}_k+\textbf{G}_k  \label{eq:ssm_1}\\ 
    \textbf{Y}_k &=\textbf{CX}_k+\textbf{V}_k ~.  \label{spm}         
\end{align}
where the first equation represents the dynamics of the system while the second represents the (sensor) observation model. Here $\textbf{X}_k \in \mathbb{R}^q$ is the state vector of the system, $\textbf{U}_k \in \mathbb{R}^p$ the input (or control) vector, and $\textbf{Y}_k \in \mathbb{R}^N$ the output vector of sensors. Matrices $\textbf{A}$, $\textbf{B}$ and $\textbf{C}$ are determined by the physics of the system as well as the sensors.  $\textbf{G}$ and $\textbf{V}$ are Gaussian white noise with zero mean and covariance matrices $\textbf{R}_\textbf{G}$ and $\textbf{R}_\textbf{V}$, respectively. $\textbf{X}_0$, $\textbf{G}_k$ and $\textbf{V}_k$ are assumed to be independent.  Assuming the noises $\textbf{G}$ and $\textbf{V}$ are small, the next system state, $\textbf{X}_{k+1}$, primarily depends on the current system state, $\textbf{X}_{k}$, and the current input $\textbf{U}_k$, while the current output of the sensors, $\textbf{Y}_k$, primarily depends on the current system state $\textbf{X}_{k}$.

The Kalman filter state estimation can be separated into two steps, a prediction step and an update step. In the prediction step, the predicted state (of time $k$ based on the value at time $k-1$), $\hat{\textbf{X}}_{k|k-1}$  and the corresponding uncertainty measure of the prediction, $\textbf{P}_{k|k-1}$ are calculated as follows:
\begin{align}
    \hat{\textbf{X}}_{k|k-1} &= \textbf{A}\hat{\textbf{X}}_{k-1|k-1} + \textbf{B} \textbf{U}_{k}\\
    \textbf{P}_{k|k-1} &= \textbf{A} \textbf{P}_{k-1|k-1} \textbf{A}^{\text{T}} + \textbf{R}_\textbf{W} ~,
\end{align}
Upon observing a measurement $\textbf{Y}_k$, the estimated state and uncertainty measure are updated as follows:
\begin{align}
    \textbf{K}_k &= \textbf{P}_{k|k-1}\textbf{C}^\text{T}(\textbf{C} \textbf{P}_{k|k-1} \textbf{C}^\text{T} + \textbf{R})^{-1}\\
    \hat{\textbf{X}}_{k|k} &= \hat{\textbf{X}}_{k|k-1} + \textbf{K}_k(\textbf{Y}_k - \textbf{C}\hat{\textbf{X}}_{k|k-1})\\
    \textbf{P}_{k|k} &= (\textbf{I} - \textbf{K}_k \textbf{C}) \textbf{P}_{k|k-1} ~,
\end{align}
where the updated state, $\hat{\textbf{X}}_{k|k}$, is a weighted sum of the estimated state and the innovation ($\textbf{Y}_k - \textbf{C}\hat{\textbf{X}}_{k|k-1}$). The weight  depends on the uncertainty measure $\textbf{P}_{k|k-1}$: the more uncertain the estimated state is, the more weight is placed on the new observation.

The group testing method requires the fault detection method to identify whether an arbitrary group of sensors contains any faulty member. The idea of using Kalman filtering for group testing lies in its ability to estimate the state of the underlying system from the observations of arbitrary sets of sensors. For example, if one wants to estimate the system state by using the outputs from sensors 1, 3 and 4, the observation model (Eq. (\ref{spm})) can be changed to $\textbf{Y}_k^\prime =\textbf{C}^\prime\textbf{X}_k+\textbf{V}_k^\prime$, where $\textbf{Y}_k^\prime$ ($\textbf{V}_k^\prime$) contains only the $1^{st}$ and $3^{rd}$ components of $\textbf{Y}_k$ ($\textbf{V}_k$) and $\textbf{C}^\prime$ contains the $1^{st}$ and $3^{rd}$ rows of $\textbf{C}$. 
The dynamic equation of the system (Eq. (\ref{eq:ssm_1})) remains the same.
With this in mind, after selecting a test group of sensors $\Phi_i$, we can split it into two subgroups $A$ and $B$, and use the observations from each subset to separately estimate the state of the underlying system (thus it is required that the test group $\Phi_i$ contain at least two sensors), and check whether the two are consistent.

\begin{figure*}[tb]
\centering
\includegraphics[width=15cm]{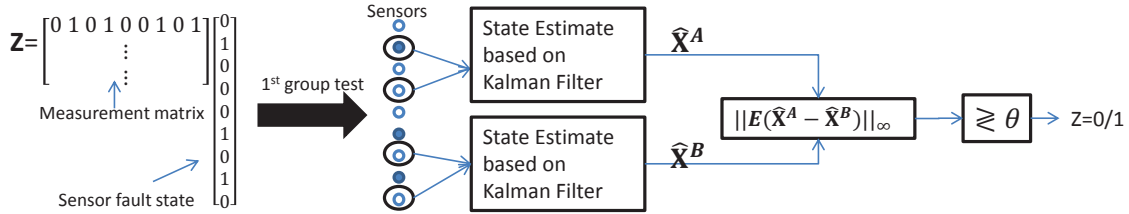}
\caption{State diagram of the proposed sensor fault detection method.}
\label{fig:overview}
\end{figure*}

Specifically, denote the estimated states of the system computed from observations of the subgroups A and B as $\hat{\textbf{X}}^{A}_{k|k-1}$ and $\hat{\textbf{X}}^{B}_{k|k-1}$, respectively. The difference between the two estimated states is given by:
\begin{align}
    \textbf{e}_k &= \hat{\textbf{X}}^{A}_{k|k-1} - \hat{\textbf{X}}^{B}_{k|k-1}~.   
\end{align}
As all states estimated from the Kalman filter are unbiased (i.e., $E[\hat{\textbf{X}}_{k|k-1}]={\textbf{X}}_{k}$) \cite{maybeck1979stochastic}, the expected difference $E[\textbf{e}_k]=E[\hat{\textbf{X}}^{A}_{k|k-1}]-E[\hat{\textbf{X}}^{B}_{k|k-1}]=0$ if neither $A$ nor $B$ contains any faulty sensor. 
Otherwise this expectation is non-zero.
Therefore, a threshold can be used 
to decide whether a group of sensors $\Phi_i$ contains any faulty sensors: if the difference between the two state estimates 
is larger than this threshold, then $\Phi_i$ is regarded as having at least one faulty sensor and the corresponding entry in $Z$ will be set to $1$; otherwise the corresponding entry in $Z$ is set to $0$. Fig. \ref{fig:overview} gives an overview of this approach.


After obtaining the group test result $Z$, the sensor fault state is recovered by a straightforward maximum likelihood (ML) decoding. The recovery algorithm evaluates all ${N\choose d}$ possible fault states and chooses the one such that the group testing result $Z$ is most likely, i.e., choose $\nu^\star$ if
\begin{equation}
    P(Z | L_\nu^\star) > P(Z | L_\nu) \quad \forall \nu \neq \nu^\star \label{ML}
\end{equation}
where $L_\nu$ denotes any possible fault state and $\nu \in \{1,2, \ldots,\sum_0^d{N\choose d}\}$.
However, the probability measure in Eq. (\ref{ML}) may be difficult to obtain; in our case it depends on the threshold used in group testing.  We will thus simply assume that each group test has the same false positive and false negative probabilities and use minimum distance decoding.  For each possible fault state $L_\nu$, the recovery algorithm calculates the Hamming distance, defined as the number of distinct entries, between the predicted output $\Phi L_\nu$ and the detection outcome $Z$. Fault states with smaller Hamming distance is preferred.  Among fault states having the same Hamming distance from $Z$, states with a smaller support are preferred as the probability of a sensor being faulty is $<1/2$.  If this still results in a tie, then the recovery algorithm will choose randomly.


\section{A Combinatorial Group Testing Based Fault Detection Method}
\label{sec:methodology}

In this section, a Combinatorial Group testing (CGT) based fault detection method is presented \cite{lo2013efficient}. This section focuses on the design of the test groups or measurement matrix $\Phi$. The group test is preformed using Kalman filtering as described in Section \ref{sec:KF}.  Consider a network of $N$ sensors monitoring an underlying physical system that can be modeled as a linear dynamical system.  Assume any sensor in the network can be faulty and that at most $d$ of them are faulty at any given time.  The dynamic evolution of the underlying system as well as observations by the sensors can be expressed similarly as in (\ref{spm}):
\begin{align}
    \textbf{Y}_k &=\textbf{CX}_k+\textbf{V}_k+\textbf{E}_k,  \label{spm2}         
\end{align}
where the additional vector $\textbf{E}_k$ is an unknown error vector induced by sensor faults: its $i^{th}$ component is zero if sensor $i$ is not faulty.


%


\subsection{Group Selection and the Number of Group Tests}
\label{sec:group_selection}

Recall the fault detection problem represented as $Z= \Phi S$, where S represents the fault state of sensors (``1'' means faulty).
%
As the detection performance largely depends on $\Phi$, our primary task is in determining the entries of $\Phi$, i.e., which sensors include in each test.  In this sub-section we focus  on the non-adaptive CGT method, whereby $\Phi$ is designed prior to the tests.  

\rev{A common way of selecting test groups, which we adopt in this study,} is to design a
disjunct measurement matrix. A $d$-disjunct matrix has the property that for any $d+1$ columns, there is always a row with entry 1 in a column and zeros in all the other $d$ columns. For instance, the measurement matrix in Example \ref{exam:ex1} is 1-disjunct (since any two columns differ in at least one row) but is not 2-disjunct. \rev{The reason a $d$-disjunct matrix is desirable, especially in the case when group tests are error-free, is because its output vector $Z$ is distinct for different $d$-sparse vectors $S$ (a vector is $d$-sparse if it has at most $d$ non-zero entries), which means that the exact recovery of a $d$-sparse fault state vector $S$ is guaranteed with a $d$-disjunct $\Phi$.}
One simple method to generate a $d$-disjunct measurement matrix $\Phi$ with high probability is to generate each entry randomly such that $\Phi(i,j)=1$ has probability $1/2$.

The quality of a measurement matrix is reflected in the number of tests needed (the number of rows in $\Phi$) to gain enough information in order to correctly recover the fault state $S$.  If the group tests are error free and the faulty sensors are distributed uniformly at random, then the necessary and sufficient number of rows in $\Phi$ are $\Omega(d\log(N/d))$ and $\mathcal{O}(d\log(N))$, respectively \cite{Gilbert_Recovering_simple_signals}.  Under the worse-case distribution of faults (i.e., adversarial fault model), the necessary and sufficient number of rows in $\Phi$ are $\Omega(\frac{d^2\log(N)}{\log(d)})$ and $\mathcal{O}(d^2\log(N))$, respectively \cite{Gilbert_Recovering_simple_signals}.
%
%
%

The group tests in our problem is not error-free since detection using Kalman filtering is inherently noisy.  Noisy group testing problems are relatively less studied than their noise-free counterpart.
A recent study \cite{Atia:2012:BooleanCS_meansurements} evaluated the number of tests required for two noisy group testing scenarios: 1) Additive model, where a negative group test result may turn to positive with certain probability; 
and 2) Dilution model, where a faulty sensor may act normal (diluted) with certain probability in a group test. The sufficient number of tests for the additive model and dilution model, under the worst-case distribution of faults, are shown to be $\mathcal{O}({d^2\log(N)}/{(1-\alpha)})$ and $\mathcal{O}({d^2\log(N)}/{(1-\beta)^2})$, respectively. However, for group tests that can have both false alarm and miss detection, as in our case, the required number of tests remains an open question.




\subsection{Practical Implementation}
The method outlined above can be implemented in two ways. The first is as a post processing of data already collected at a cluster head or central location. 
The second is in a form of real-time sequential process, where a control center solicits input from a single group of sensors at a time.  A single group test is then performed over this group of input.  This is followed by soliciting input from the next group, and so on.  Note that as long as the fault state of the underlying system remains unchanged, the fault state estimate can be done over different segments of observations over time.  In other words, the data provided by each group need not be synchronized and can be generated on demand.

\section{Bayesian Group Testing}
\label{sec:BGT}
We next present a novel adaptive group testing method based on Bayesian inference. The combinatorial group testing method presented in the previous section designs the entire set of tests (i.e., the entire $\Phi$) before carrying out any group test. The result of each group test, however, may provide valuable information on the sensor state.  For instance, in the extreme case when group tests are error-free, a negative result implies that all items in that test are normal; thus no further test is required for these items.  By taking previous test results into account (i.e., adapt to the group test results), the sensor state may be identified with fewer number of tests compared to the combinatorial group testing method. This idea was adopted in several studies \cite{hwang1972binary,allemann2003BGT_split,du1993combinatorial}.  

Our method maintains a probability measure on the sensor fault state vector, which is updated following each group test using Bayesian inference.  The updated state estimate is then used to determine the next test pool.  This process is repeated until the change in the state estimates is sufficiently small. \rev{As we shall see, compared to existing adaptive group testing methods, our algorithm is designed specifically for noisy group tests so that errors do not propagate.}

In the following presentation, subscript $k$ is used to denote the $k^{th}$ component (row) of a vector (matrix) and superscript $k$ to denote the collection of a variable from time $1$ to $k$. Specifically, denote by $\Phi^k=\{\Phi_1, \Phi_2, \ldots, \Phi_k\}$ the set of tests used up to time $k$, where $\Phi_k$ is the $k^{th}$ row vector of $\Phi$, 
and $Z^k=\{Z_1, Z_2, \ldots, Z_k\}$ the set of test results up to time $k$. Let ${\cal S}$ be the collection of all possible sensor fault states 
$\{S=(S_1, S_2, \ldots, S_N): S_i \in \{0,1\}\})$.  We define two probability measures. 
The first is $P_{S,k}=P(S|\Phi^k, Z^k)$, the probability of the sensor state being $S\in {\cal S}$ after the $k^{th}$ test; the second is $P_{i,k}$, the probability of sensor $i$ being normal after the $k^{th}$ group test. By definition, we have
$P_{i,k}=\sum_{S\in {\cal S}:S_i=0}P_{S,k}$.

For the $(k+1)^{th}$ test $\Phi_{k+1}$, it is desirable to select sensors such that the test result $Z_{k+1}$ provides the most information for the estimation of the true sensor state. Basic information theory result \cite{cover2012elements} tells us that maximizing the information content is equivalent to maximizing the variance of $Z_{k+1}$. This criterion can be expressed as follows:
\begin{align}
\label{eq:max_var}
    \Phi_{k+1}^\ast 
                    &=\argmax_{\Phi_{k+1}} VAR[Z_{k+1}|\Phi_{k+1}, \{P_{S,k}\}_{S\in {\cal S}}] ~.
\end{align}
$Z_{k+1}$ conditioned on $\Phi_{k+1}, \{P_{S,k}\}_{S \in {\cal S}}$
has a Bernoulli distribution.  If we denote by $\Omega_k$ the probability that all sensors in test pool $\Phi_{k+1}$ are normal given the estimate after the $k^{th}$ observation, then the above variance is given as follows, noting that $Z_{k+1}=0$ either when all sensors in $\Phi_{k+1}$ are normal and the group test is correct or when at least one sensor in $\Phi_{k+1}$ is abnormal and the group test is incorrect, i.e., $Z_{k+1}=0$ with probability $((1-\alpha)\Omega_k+\beta(1-\Omega_k))$, and similarly $Z_{k+1}=1$ with probability $(\alpha \Omega_k+(1-\beta)(1-\Omega_k))$.
\begin{align}
    &VAR[Z_{k+1}|\Phi_{k+1}, \{P_{S,k}\}_{S\in {\cal S}}]  \nonumber \\
    &=((1-\alpha)\Omega_k+\beta(1-\Omega_k))(\alpha \Omega_k+(1-\beta)(1-\Omega_k)) \nonumber \\ 
    &=\beta-\beta^2+(1-2\beta)(1-\alpha-\beta)\Omega_k-(1-\alpha-\beta)^2\Omega_k^2 \label{eq:var_Z_final}~.
\end{align}
The above computation, however, is generally intractable due to the large state space ${\cal S}$ when the number of sensors is large.  We thus adopt the following  approximation by assuming conditional independence between different sensors' fault states, i.e., 
\begin{equation}
\label{eq:condi_indep}
    P(S_1,S_2, \ldots, S_N|\{P_{S,k}\}_{S \in {\cal S}})=\prod_{i\in N} P(S_i|\{P_{S,k}\}_{S \in {\cal S}})~, ~ \forall k ~.
\end{equation}
%
With this assumption we have
$\Omega_k=\prod_{i \in \Phi_{k+1}}P_{i,k}$, 
where we have used $i \in \Phi_{k+1}$ to mean that the $i^{th}$ component of $\Phi_{k+1}$ is $1$.
%

While this assumption allows us to compute (\ref{eq:var_Z_final}), finding the optimal solution to 
(\ref{eq:max_var}) 
remains hard when the number of sensors is large. 
Toward this end we propose a greedy algorithm for choosing a good $\Phi_{k+1}$ efficiently, by observing from
(\ref{eq:var_Z_final}) that its maximum is achieved 
when $\Omega_k^\ast=(1-2\beta)/(2(1-\alpha-\beta))$.
The greedy algorithm starts with a random sensor and calculates $\Omega_k$; in each successive step it selects a sensor such that the resulting new value of $\Omega_k$ is as close to $(1-2\beta)/(2(1-\alpha-\beta))$ as possible.
This is repeated until no additional sensor can bring $\Omega_k$ closer to $(1-2\beta)/(2(1-\alpha-\beta))$. As $\Omega_k$ is monotonically decreasing in the inclusion of new sensors, the algorithm is guaranteed to terminate with a new test pool.

Having designed $\Phi_{k+1}$ and observed $Z_{k+1}$, the probability $P_{S,k+1}$ can be updated from $P_{S,k}$ for all $S\in {\cal S}$: 
\begin{align}
    P_{S,k+1}&=P(S|\Phi^{k+1}, Z^{k+1})=\frac{P(S, \Phi^{k+1}, Z^{k+1})}{P(\Phi^{k+1}, Z^{k+1})} \nonumber\\
    &=\frac{P(Z^{k+1}|Z^k, S, \Phi^{k+1})P(S|\Phi^{k}, Z^{k})P(\Phi^{k+1}, Z^{k})}{P(\Phi^{k+1}, Z^{k+1})} \nonumber\\
    &= P(Z_{k+1}|Z^k, S, \Phi^{k+1})P_{S,k}/\Delta_k~, \label{eq:bayesian_update}
\end{align}
where $\Delta_k$ is the normalizing factor $P(\Phi^{k+1}, Z^{k})/P(\Phi^{k+1}, Z^{k+1})$, and is equal to $\sum_S P(Z_{k+1}|Z^k,S,\Phi^{k+1})P_{S,k}$.
Note that
$P(Z_{k+1}|Z^k, S, \Phi^{k+1})=P(Z_{k+1}|\Phi_{k+1}S)$ as $Z_{k+1}$ only depends on the error-free test result $\Phi_{k+1}S$;
recall the two type of errors are given by $P(Z_{k+1}=1|\Phi_{k+1}S=0)=\alpha$ and $P(Z_{k+1}=0|\Phi_{k+1}S=1)=\beta$. 

To update the sensor state probabilities using (\ref{eq:bayesian_update}) for each $S\in{\cal S}$ can be computationally prohibitive for large $N$ ($|{\cal S}|=2^N$).  Below we show that using the conditional independence assumption we can instead update $P_{i,k+1}$ directly
without calculating $P_{S,k+1}$, thus reducing the complexities from $\mathcal{O}(2^N)$ to $\mathcal{O}(N)$.  We first calculate the normalization constant, and then update $P_{i,k+1}$ accordingly.

Given a test pool $\Phi_{k+1}$, we will refer to the set of sensor states $\{S: \Phi_{k+1}S=1\}$ as the {\em positive set}, and 
$\{S: \Phi_{k+1}S=0\}$ as the {\em negative set}. 
Note that by definition, we have $\sum_{S: \Phi_{k+1}S=0} P_{S,k} = \Omega_k$ and $\sum_{S:\Phi_{k+1}S=1} P_{S,k} = 1-\Omega_k$.
%
By separating ${\cal S}$ into these two sets, $\Delta_k$ can be calculated as follows:
\begin{align}
     \Delta_k= &\sum_S P(Z_{k+1}|\Phi_{k+1}S)P_{S,k} \nonumber\\
     = &\sum_{S:\Phi_{k+1}S=1} P(Z_{k+1}|\Phi_{k+1}S)P_{S,k}+\sum_{S:\Phi_{k+1}S=0} P(Z_{k+1}|\Phi_{k+1}S)P_{S,k} \nonumber \\
     = & P(Z_{k+1}|\Phi_{k+1}S=1)(1-\Omega_k)-P(Z_{k+1}|\Phi_{k+1}S=0) \Omega_k \label{eqn:delta}
\end{align}

Therefore, if the test result is positive, $Z_{k+1}=1$, then $\Delta_k=(1-\beta)(1-\Omega_k)-\alpha \Omega_k$; if the test result is negative, $Z_{k+1}=0$, then $\Delta_k=(\beta)(1-\Omega_k)-(1-\alpha) \Omega_k$.

We next show how $P_{i,k+1}$ is updated.
If sensor $i\in\Phi_{k+1}$, then using (\ref{eq:bayesian_update}) we have
\begin{eqnarray}
    P_{i,k+1}&=&\sum_{S: S_i= 0}P_{S,k+1}
    =1-\sum_{S: S_i=1}P_{S,k+1} \nonumber \\
    &=&1-\sum_{S: S_i=1}P_{S,k}P(Z_{k+1}|\Phi_{k+1} S=1)/\Delta_k \nonumber \\
    &=&1-(1-P_{i,k})P(Z_{k+1}|\Phi_{k+1}S=1)/\Delta_k \nonumber \\
%
   &= & \begin{cases}
    1-(1-P_{i,k})(1-\alpha)/\Delta_k & \text{if } Z_{k+1}=1,\\
    1-(1-P_{i,k})(1-\beta)/\Delta_k & \text{if } Z_{k+1}=0.
    \end{cases}
\end{eqnarray}
If sensor $i\not\in\Phi_{k+1}$, then using (\ref{eqn:delta}) we have:
\begin{align}
    P_{i,k+1}=&\sum_{S: S_i=0}P_{S,k+1}
    = \sum_{S:S_i=0, \Phi_{k+1}S=1}P_{S,k+1}+\sum_{S:S_i=0, \Phi_{k+1}S=0}P_{S,k+1}\nonumber\\
    = &\sum_{S:S_i=0, \Phi_{k+1}S=1}P_{S,k}P(Z_{k+1}|\Phi_{k+1}S)/\Delta_k
    +\sum_{S:S_i=0, \Phi_{k+1}S=0}P_{S,k}P(Z_{k+1}|\Phi_{k+1}S)/\Delta_k \nonumber \\
   = &P_{i,k}(1-\Omega_k) P(Z_{k+1}|\Phi_{k+1}S=1)/\Delta_k+ P_{i,k}\Omega_k P(Z_{k+1}|\Phi_{k+1}S=0))/\Delta_k \nonumber \\
    = &P_{i,k} \Delta_k/\Delta_k = P_{i,k} 
\end{align}
where the fourth equality is due to the independence assumption.
%
As a result, when $i\not\in\Phi_{k+1}$, the corresponding $P_{i,k+1}$ remains unchanged. 

The above computational procedure is repeated after each test, starting from some assumed initial prior $P_{i,0}$.  After $k$ tests and given $Z^k$ and $\Phi^k$, the sensor fault state $S$ can be recovered in two ways:
(1) use the maximum a posteriori probability (MAP) estimator: $\argmax_{S}$ $P(Z^k|S, \Phi^k)P_{S,k}$, or
(2) declare the $i^{th}$ sensor faulty if $P_{i,k}<\sigma$ for some predefined threshold $\sigma$, and normal otherwise. 
While both are valid, the second method is preferred as $P_{i,k}$ is readily available from the above updating procedure, whereas the MAP estimation is computationally much more complex. The performance of these two methods is similar as we show in Section \ref{sec:BGT_results}.

\section{Experimental Setup}
The proposed CGT and BGT fault detection algorithms are evaluated using a set of measured  vibration data collected by wireless sensors from the New Carquinez Bridge in California. In this section, we first present common sensor fault types and then detail the nature of the measured data and how it is used in our evaluation.


\subsection{Sensor Fault Types} \label{sec:fault_types}

\begin{figure}[tb]
\centering
\includegraphics[width=10cm]{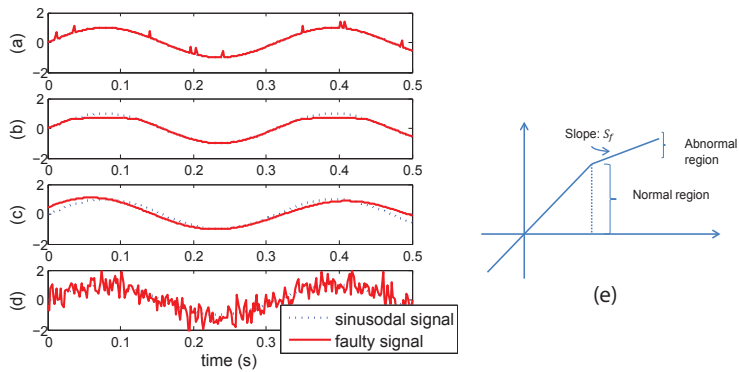}
\caption{Illustration of different faults on a sinusoidal signal: (a) Spike, (b) Non-linearity, (c) mean-drift, (d) Excessive noise and (e) non-linear fault model}
\label{fig:fault_sample}
\end{figure}

We consider four different fault types: spike, non-linear transduction, mean drift and excessive noise in the controlled experiments. These are illustrated in Fig. \ref{fig:fault_sample} on a sinusoidal signal. More specifically, a spike fault is an impulse superimposed on normal sensor measurements. They are assumed to occur randomly in time with constant or varying magnitudes (consistent with a random signal model). Moreover, the occurrence of these spikes is assumed sparse. A non-linearity fault represents an abnormal discrepancy between the sensor input and output. This fault usually happens when the measurement falls outside a certain dynamic range. In this study, a simple non-linear fault model is used as shown in Fig. \ref{fig:fault_sample}(e): when the measurement is within the normal region, the sensor output reflects the measurement; otherwise the output follows the slope $S_{f}$. A mean drift fault preserves the output dynamics but not its mean value.  This type of fault generates outputs whose mean drifts away from the true mean of the signal slowly compared to the output dynamics. Finally, excessive noise refers to a large amount of Gaussian noise in the output of a sensor. Compare to regular measurement noise, this fault has much higher amplitude such that the output signal is highly corrupted.  Note that only the non-linearity fault is a function of the measured signal while the other fault types are not.


\subsection{Bridge Vibration Data and State Estimation}
\begin{figure}[tb]
\centering
\includegraphics[width=8.5cm]{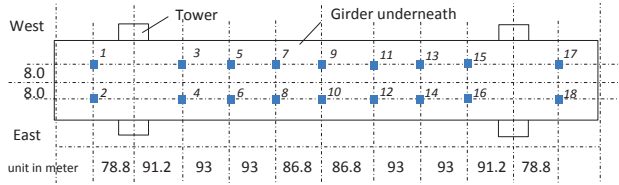}
\caption{Plan map of the deployed sensors. Credit: Yilan Zhang}
\label{fig:sensor_map}
\end{figure}

\begin{figure}[tb]
\centering
\includegraphics[width=7cm]{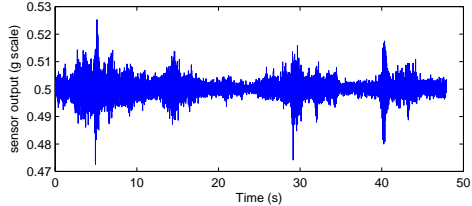}
\caption{Vibration measurement of a sensor}
\label{fig:sensor_sample}
\end{figure}

We evaluate our detection method using bridge vibration data collected by a network of 18  vibration sensors deployed on the New Carquinez Bridge in California. This is a 1056-meter long suspension bridge which connects Crockett and Vallejo. The locations of these 18 sensors are shown in Fig. \ref{fig:sensor_map}. They monitor the bridge vibration in the direction perpendicular to the bridge surface. Fig. \ref{fig:sensor_sample} shows an example of the output of a sensor when vehicles pass through. We took 18 data traces at the beginning of the deployment and performed manual inspection. \rev{Each data trace consists of 50 seconds of data sampled at 200Hz.} All tests, including  spectrum analysis and mode-shape calculation on the data suggest that the data traces are correct.





\rev{Our first task is to use the collected data to train the linear dynamical model needed in the group testing algorithms.}
For this we adopt a commonly used approach, the subspace method \cite{katayama2005subspace} which utilizes measured output (and input, if available) to calculate  model parameters such as matrices $\textbf{A}$, ($\textbf{B}$ if input data is available) and $\textbf{C}$ in the state-space model (\ref{spm2}). Notice that the excitation/input to the bridge is in general unavailable.  While input is not necessary for learning the system model by the subspace method, prior study suggests the input can be assumed to be Gaussian for large structures with complex excitations, and that this leads to a better learned system model in terms of output prediction \cite{tong1998multichannel}.
\rev{For our study,
we use half of of the vibration data from each of the 18 traces for training of the bridge dynamical model, and the other half for evaluating the group testing method.} The order of the dynamical model is set to 162 (An earlier study of the bridge, \cite{kurata2012NCQ}, indicates that a 162-order state space model is sufficient to capture the bridge dynamics), i.e., the length of the state vector is 162. The excitation inputs are assumed to be 18 degree-of-freedom Gaussian signals and each degree-of-freedom input has zero mean and variance equal to the variance of the output of the sensors.

Two experiments are then conducted to evaluate the performance of the proposed algorithms. The first is a control experiment, 
\rev{whereby different fault types are artificially created} and 
superimposed over a random subset of the data traces. The resulting data are then used for evaluation purposes.
%
Specifically, we add different types of faults to the bridge data by randomly selecting up to two sensors (a number $\varsigma$ is first chosen uniformly from $\{0,1,2\}$, and then $\varsigma$ number of intended faulty sensors are chosen uniformly among the 18 sensors).  We set the maximum number of faulty sensor to be 2 ($d=2$) so as to keep the  percentage of faulty sensors around $10\%$.  A total of 100 random runs are conducted (over the choice of the number and identity of the faulty sensors, as well as over the random injection of faults and the generation of the $\Phi$ matrix) for experiments.

In addition to the control experiment, we also evaluated the CGT and the BGT algorithms   on real sensor faults. Several weeks after deployment, sensor 11 started to exhibit errors in its data (this is again done by manual and visual inspection). As shown in Fig. \ref{fig:faulty_sensor_example}, the output of sensor 11 shows prominent spikes beyond normal fluctuation, and possibly has a shift in the mean amplitude and a small mean-drift error as well.  It should be noted that this observation is not the exact ground truth but is the closest one could get under the circumstances (the alternative is to take the sensor off the bridge and calibrate it in a lab; even if we could do so the result is only valid if the same type of faults persists in the lab setting).

\section{Performance of the Combinatorial Group Testing (CGT) Method}
\label{sec:CGT_result}
\begin{figure}[tb]
\centering
\includegraphics[width=7.5cm]{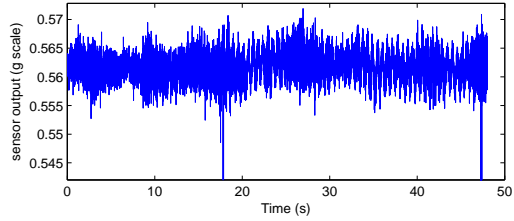}
\caption{Abnormal vibration measurement of sensor 11.}
\label{fig:faulty_sensor_example}
\end{figure}

In this section we evaluate the performance of the CGT algorithm. 
The performance in detecting different fault types is evaluated by control experiments. The algorithm is then evaluated on detecting the real faulty sensor shown in Fig.~\ref{fig:faulty_sensor_example}. Finally, we compare the CGT algorithm to non-group testing methods, in terms of accuracy and efficiency.


%
%

\begin{figure}[ht]
\centering
\begin{minipage}[b]{0.45\linewidth}
\includegraphics[width=7.5cm]{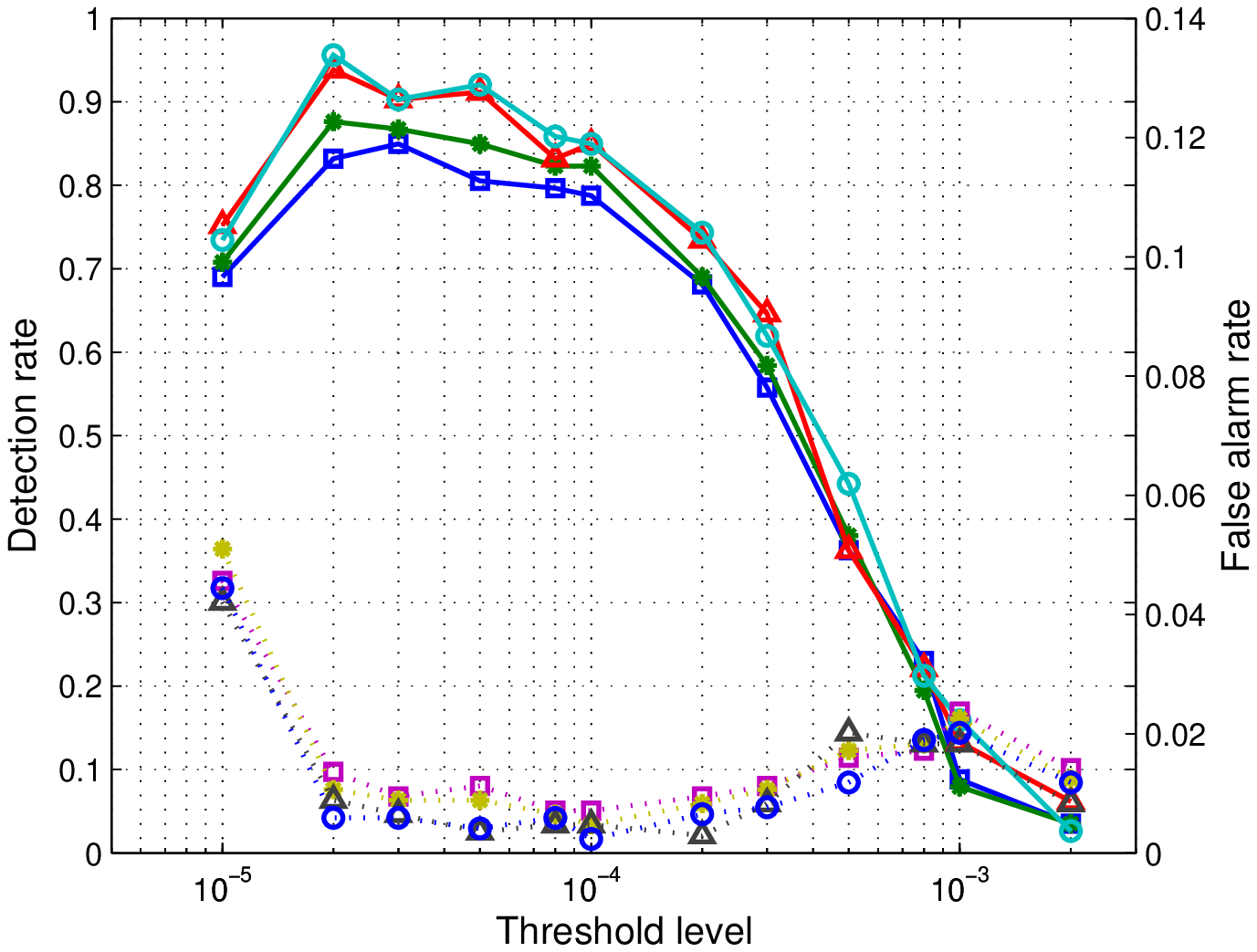}
\caption{Detection and false alarm rated for spike faults.}
\label{fig:spike_005_1}
\end{minipage}
\quad
\begin{minipage}[b]{0.45\linewidth}
\includegraphics[width=7.5cm]{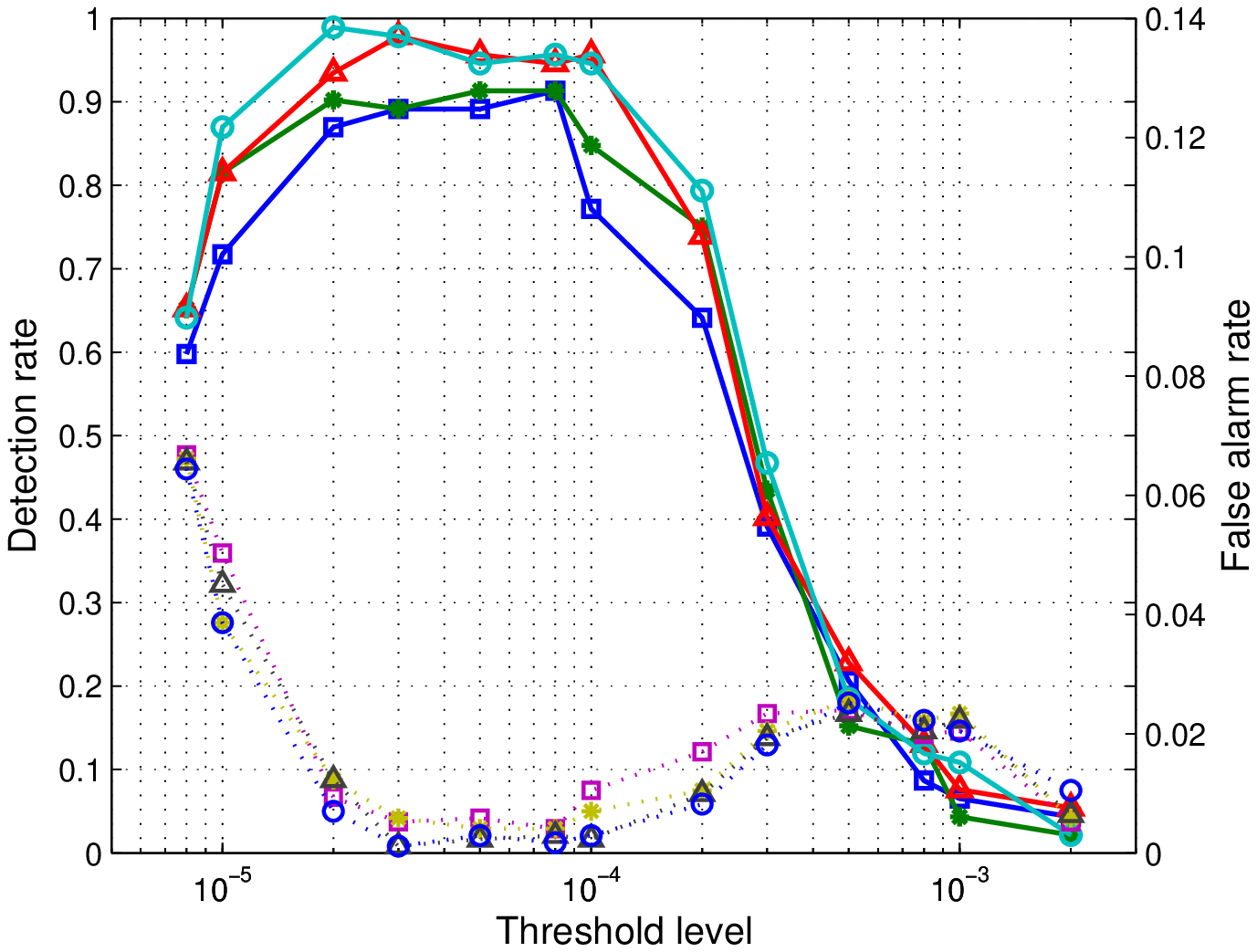}
\caption{Detection and false alarm rates for non-linearity faults.}
\label{fig:non-lin_08_03}
\end{minipage}
\end{figure}

\begin{figure}[ht]
\centering
\begin{minipage}[b]{0.45\linewidth}
\includegraphics[width=7.5cm]{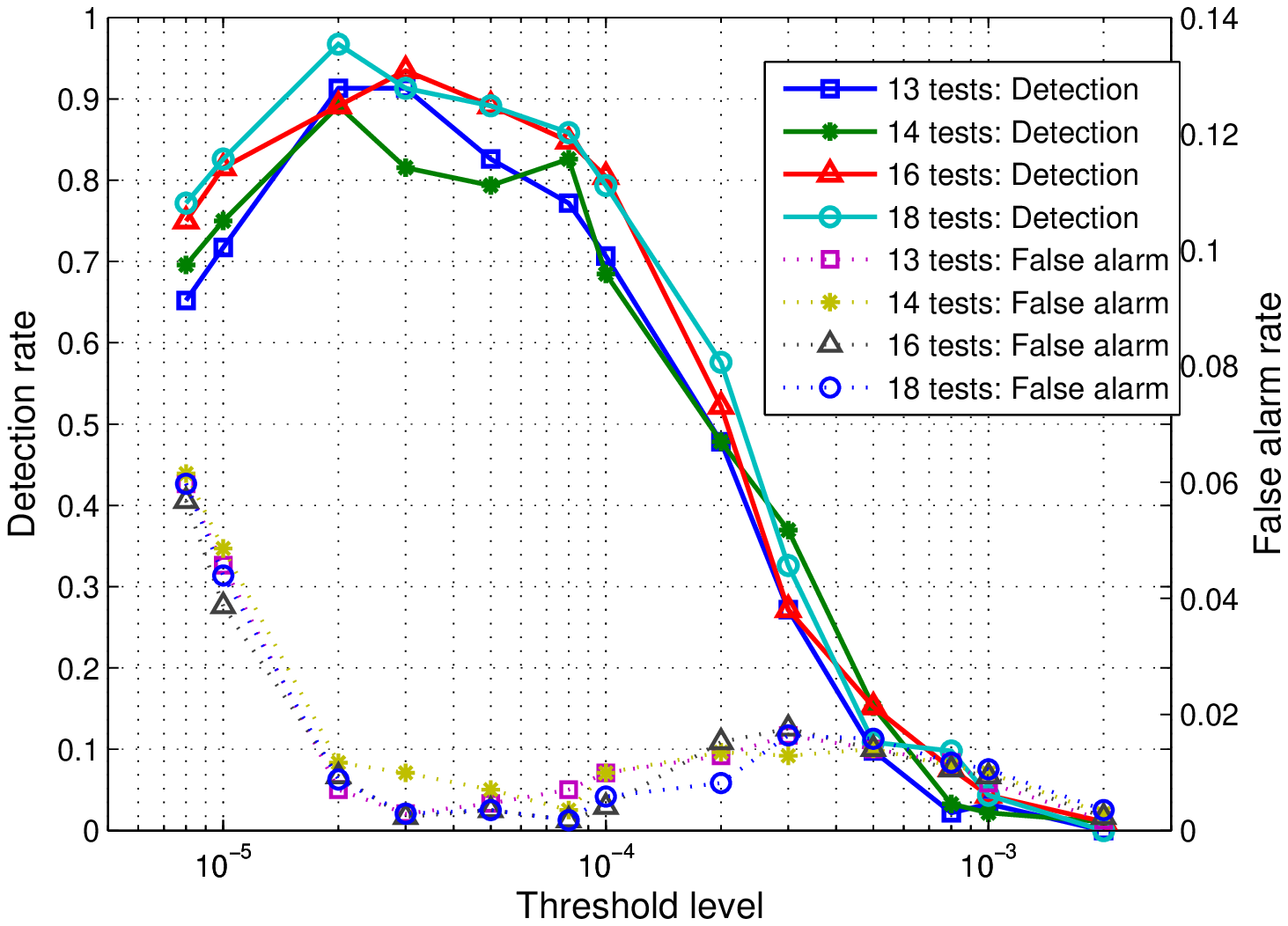}
\caption{Detection and false alarm rates for mean-drift faults.}
\label{fig:Mean_Drift_5_05}
\end{minipage}
\quad
\begin{minipage}[b]{0.45\linewidth}
\includegraphics[width=7.5cm]{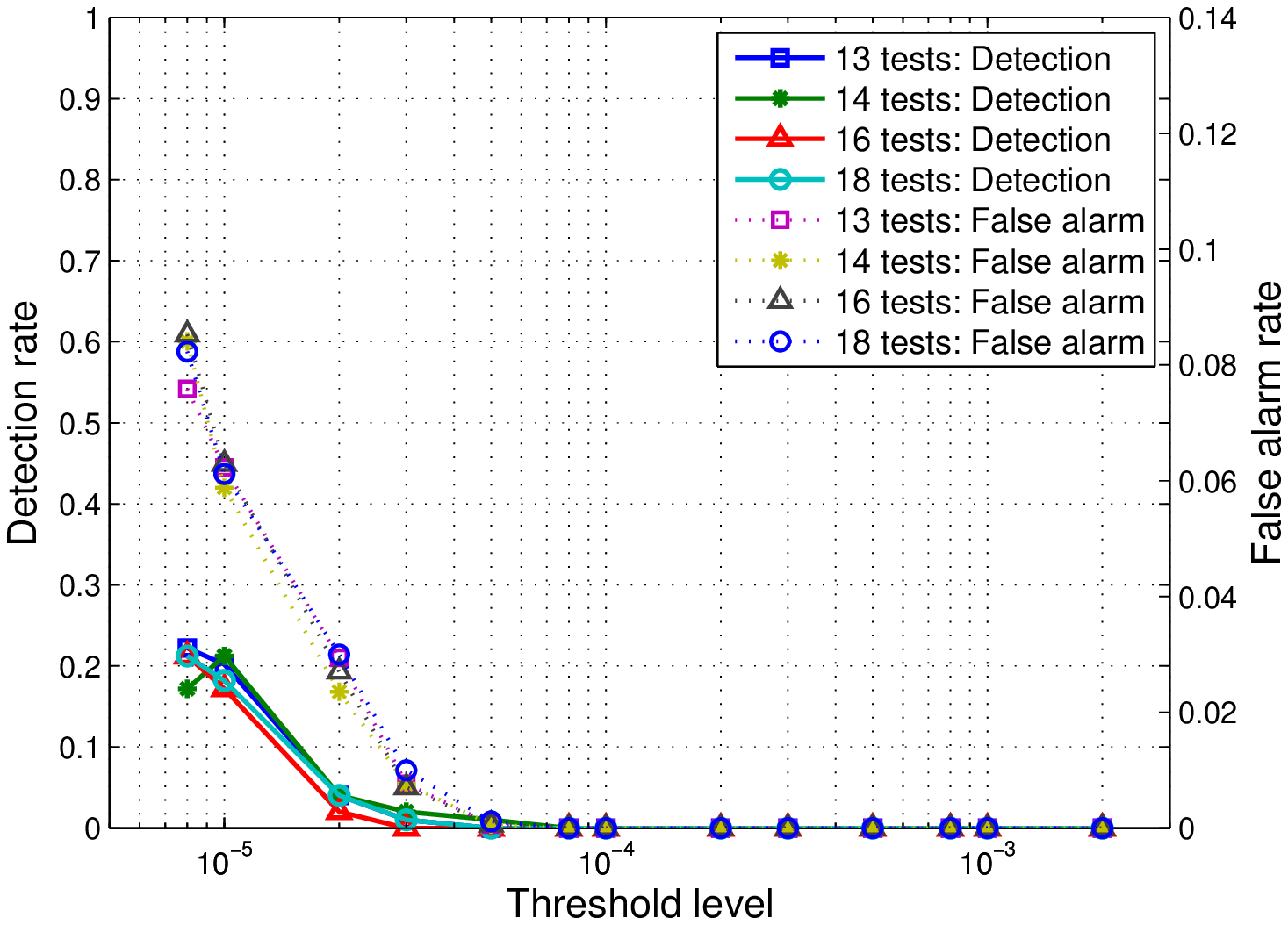}
\caption{Detection and false alarm rate for excessive noise.}
\label{fig:Gaussian_05}
\end{minipage}
\end{figure}

We first examine the performance of the CGT algorithm as a function of the detection threshold used in each group test and the number of tests performed.
Fig. \ref{fig:spike_005_1} shows the detection rate (the number of detected faulty sensors over the total number of faulty sensors) and false alarm rate in detecting spike faults under different number of tests and threshold levels.
The spike fault was set to appear at $5\%$ of the samples and have mean amplitude equal to the variance of the sensor output, which is common among spike faults in sensors.
As can be seen, as the number of tests increases, the detection rate increases while false alarm decreases. When 14 tests are used, the detection rate is above $85\%$ and false alarm  is below $1\%$, with a threshold of $2\times 10^{-5}$.  Similarly, when 16 tests are used, the accuracy is over $93\%$ and remains above $80\%$ with a threshold less than $2\times 10^{-4}$.

In all cases we see a fairly wide region of threshold values within which the method enjoys high detection rate ($>80\%$) and low false alarm ($<2\%$).  This is clearly a desired operating regime for this method.  We observe that the detection rate first and then drops slowly with increase in the threshold. When the threshold increases beyond a certain value (e.g., $3\times 10^{-4}$), the detection rate quickly drops and eventually reaches zero.  The false alarm moves in the opposite direction though to a lesser degree. To explain this phenomenon we note there are two sources of error at play, one due to Kalman filtering and the other due to the recovery algorithm.  When the threshold is very low, measurement noise or inaccuracy in the model could easily result in false positive in the the group test. These incorrect group test results cause the recovery algorithm to err, thus lead to both high false alarm and low detection rate. As the threshold increases the error from recovery decreases, which more than compensates for the decreased sensitivity in the group tests, achieving an overall better tradeoff.  When the threshold increases beyond a certain level, the group test becomes insensitive to faults and eventually declares all groups normal, resulting in reduced detection rate and false alarm.

The same evaluation is done for the other fault types; these are shown in Figs. \ref{fig:non-lin_08_03} and \ref{fig:Mean_Drift_5_05}.
For the non-linearity fault shown in Fig. \ref{fig:non-lin_08_03}, 
the normal dynamic range is set to $80\%$ of the output maximum, with a slope in the abnormal region of $0.3$.  For the mean-drift error shown in Fig. \ref{fig:Mean_Drift_5_05}, the mean-drift has a maximum frequency of $5Hz$ and a magnitude of $50\%$ of the sensor output variance. All these results show similar behavior to those observed in the spike fault case. Within the preferred threshold range, the detection rate generally exceeds $80\%$ in accuracy while false alarm remains low. Furthermore, the preferred threshold range is smaller when the fault is less pronounced.
Finally, the detection performance is tested when the sensor data is corrupted by excessive Gaussian noise with zero mean and variance equal to $50\%$ of the variance of sensor output. The result presented in Fig. \ref{fig:Gaussian_05} shows that the proposed method is not recommended for detecting this type of fault. The poor performance in this case is due to the fact that Kalman filtering, in computing statistically optimal estimates of the system state, tends to eliminate noise variance existing in the sensor measurement.  Consequently, zero-mean noise is sufficiently suppressed in the estimate and may not be  reflected in the residual of a group test.



For detecting the faulty sensor 11 shown in Fig. \ref{fig:faulty_sensor_example}, we used our algorithm on the 18 sensors with 6 and 8 tests respectively. Under the same preferred threshold range (between $3\times 10^{-3}$ and $1\times 10^{-4}$) shown in the control experiment, our algorithm was able to identify the faults in sensor 11, with a detection rate $>78\%$ ($>92\%$) and false alarm $<1.8\%$ ($<0.7\%$) when using 6 (resp. 8) tests.

Next, the proposed combinatorial group-testing based detection method is compared to two existing Kalman-filter based methods: Kobayashi et al. \cite{kobayashi2003application} and Da et al. \cite{dalin:kalman}. 
Both Kobayashi and Da are based on a bank of Kalman filters. Specifically, with $N$ sensors in the network, $N$ fault detection tests ($N$ Kalman filters) are required to evaluate all sensors in the network. In each test, all sensors but one are involved, i.e., test $i$ uses $N-1$ sensors excluding sensor $i$.  A key assumption in this method is that there is {\em only one} faulty sensor in the network, thus the test which does not contain the faulty sensor will have different characteristics than the other $N-1$ tests, and thus the single faulty sensor may be identified.

The difference between these two methods lies in how to compare the test outcomes to determine the different characteristics with and without the faulty sensor. Under the method by Kobayashi, the estimated sensor output from the Kalman filter is compared to the corresponding observed sensor output. The test which does not contain the faulty sensor will have higher consistency result than the other tests. \rev{Under the method by Da, a reference system state estimate is generated by using all $N$ sensor inputs, to which each test compares the estimated system state (from $N-1$ sensors).} 
The test that does not contain the faulty sensor is supposed to have lower consistency result because the reference contains the faulty sensor while the test does not.

\begin{figure*}[tb]
\centering
\includegraphics[width=16.5cm]{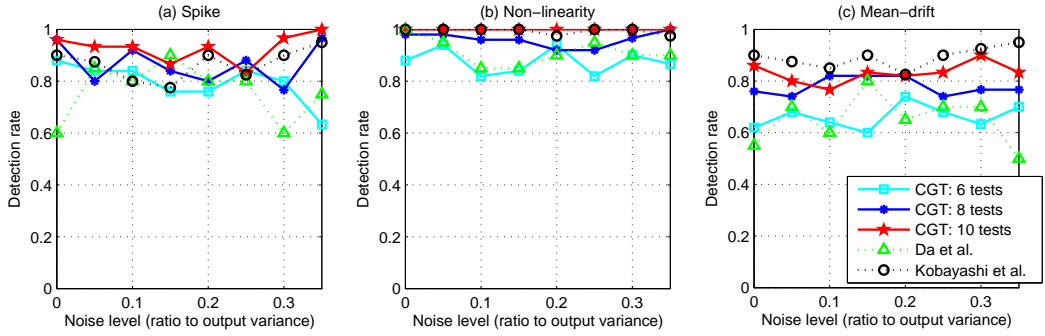}
\caption{Detection rate under different measurement noises and fault types with a single faulty sensor.} 
\label{fig:methodsVSnoise_nonadapt}
\end{figure*}

\begin{figure*}[tb]
\centering
\includegraphics[width=16cm]{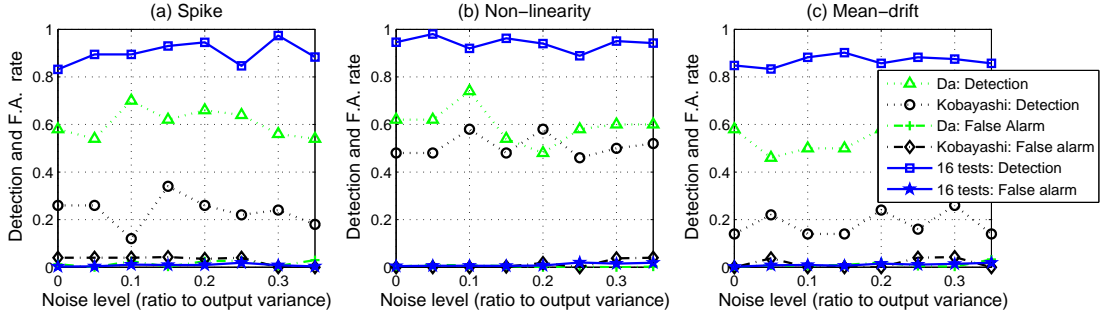}
\caption{Detection rate under different measurement noises and fault types with two faulty sensors.}
\label{fig:compare_2sensors}
\end{figure*}

Fig. \ref{fig:methodsVSnoise_nonadapt} shows the detection rate of the three methods under different types of faults, different measurement noises, and with a single faulty sensor, using the same set of bridge data. 
As we can see, Kobayashi and Da's methods achieve similar performance as our proposed method when 8 to 10 tests are used. This result is to be expected when the assumption of no more than one faulty sensor holds, since all methods are based on Kalman filtering. As shown in Section \ref{sec:preliminaries}, the the complexity of Kalman filtering largely depends on the size of the system state $\textbf{s}$, rather than the number of sensors used in state estimation. \emph{One} detection test of Da's and Kobayashi's algorithms has similar complexity as \emph{one} group test of the proposed method if the sensor network size remains the same. Therefore, our proposed method is able to achieve similar, and sometimes better, accuracy when around 8 to 10 tests are used, which is about half of the complexity of Kobayashi's and Da's method (18 tests). The results in Fig. \ref{fig:methodsVSnoise_nonadapt} also suggest, as seen earlier, that Kalman filter based fault detection systems are insensitive to Gaussian measurement noise. 
No significant degradation in the detection rate and false alarm is observed when the variance of the measurement noise increases from 0\% of output variance to 30\% of output variance.

When the system has two faulty sensors, the performance of Kobayashi and Da's methods deteriorates sharply as the reference systems are contaminated by faulty sensor observations. If the false alarm rate is restricted to a reasonable level (5\%), the accuracy of Da's method drops to about 55\% and Kobayashi's method drops to about 50\% for non-linearity fault and to about 20\% for spike and mean drift faults (Fig. \ref{fig:compare_2sensors}). At the same time, the proposed algorithm maintains over 85\% of accuracy for all fault types. Therefore, compared to other model-based methods, the proposed CGT method has fewer assumptions on the underlying system and the nature of the faults. It achieves high accuracy with much lower complexity than existing methods, which is crucial for very large sensor networks. Furthermore, the above comparison shows that the proposed method is insensitive to measurement noise.  When the system has three faulty sensors, the CGT method is able to achieve high detection rate ($90\%$ or higher) by increasing the number of tests (25 tests for detecting spike, 24 tests for detecting non-linearity and 27 tests to detect mean-drift). \rev{While these exceed the size of the network (18 sensors), this method does not require the existence of a reference system/sensor.}

\section{Performance of the Bayesian Group Testing (BGT) Method}
\label{sec:BGT_results}
The fault detection performance of the BGT method is evaluated by two experiments. Under the first experiment we use the same bridge sensor data as we did for CGT. Hence, the CGT and the BGT methods can be directly compared. The second experiment uses a much larger system (with 1000 sensors).  In this context the BGT method is compared to a well-known divide-and-conquer based adaptive group testing method proposed by Hwang \cite{hwang1972binary}. The impact of the initial prior, $P_{i,0}$, on the fault detection performance is also addressed.

\subsection{Performance of BGT on the New Carquinez Bridge sensors}

\begin{figure}[ht]
\centering
\begin{minipage}[b]{0.45\linewidth}
\includegraphics[width=7.5cm]{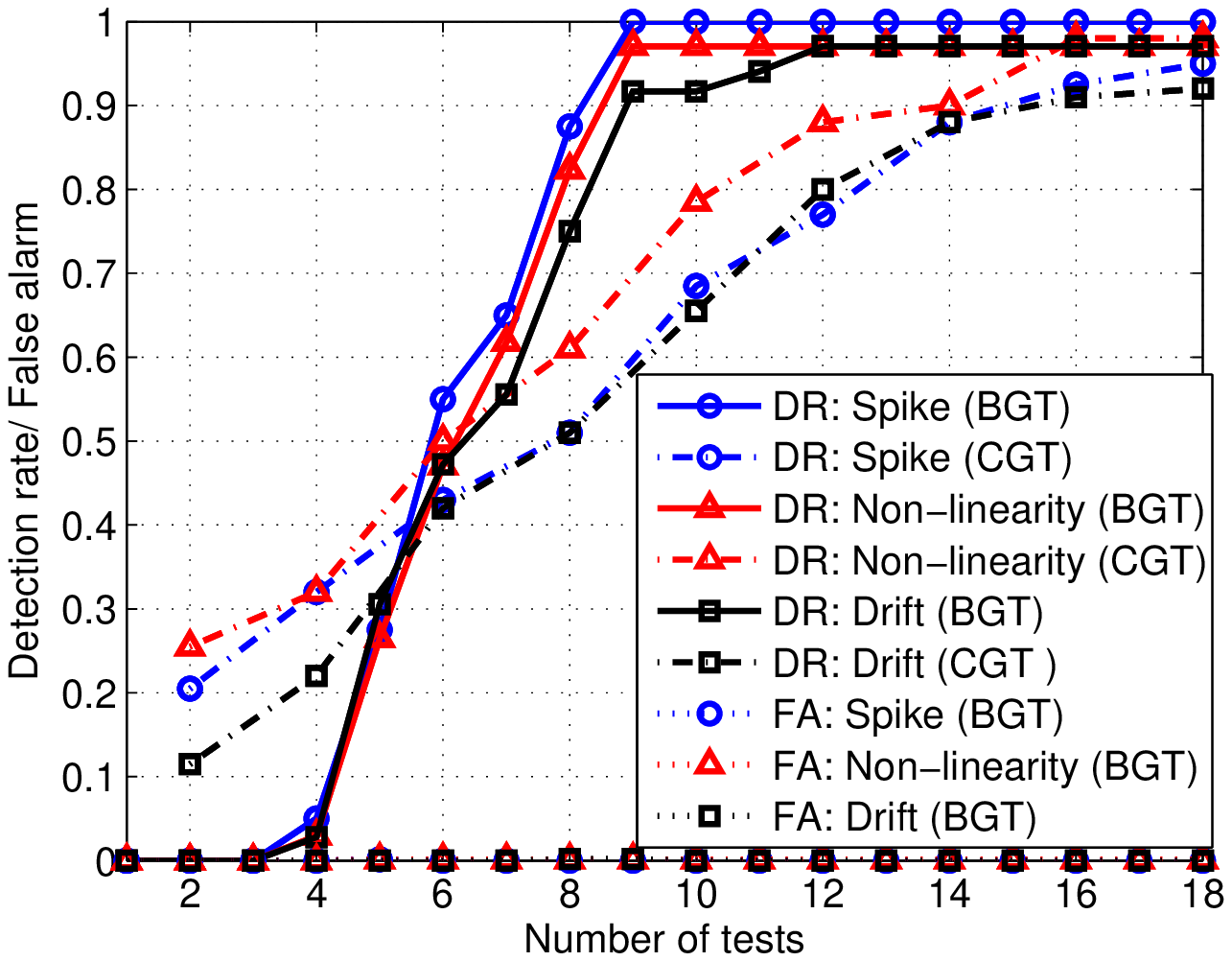}
\caption{The fault detection performance of the CGT method and the BGT method.}
\label{fig:BGT_VS_CGT}
\end{minipage}
\quad
\begin{minipage}[b]{0.45\linewidth}
\includegraphics[width=7.5cm]{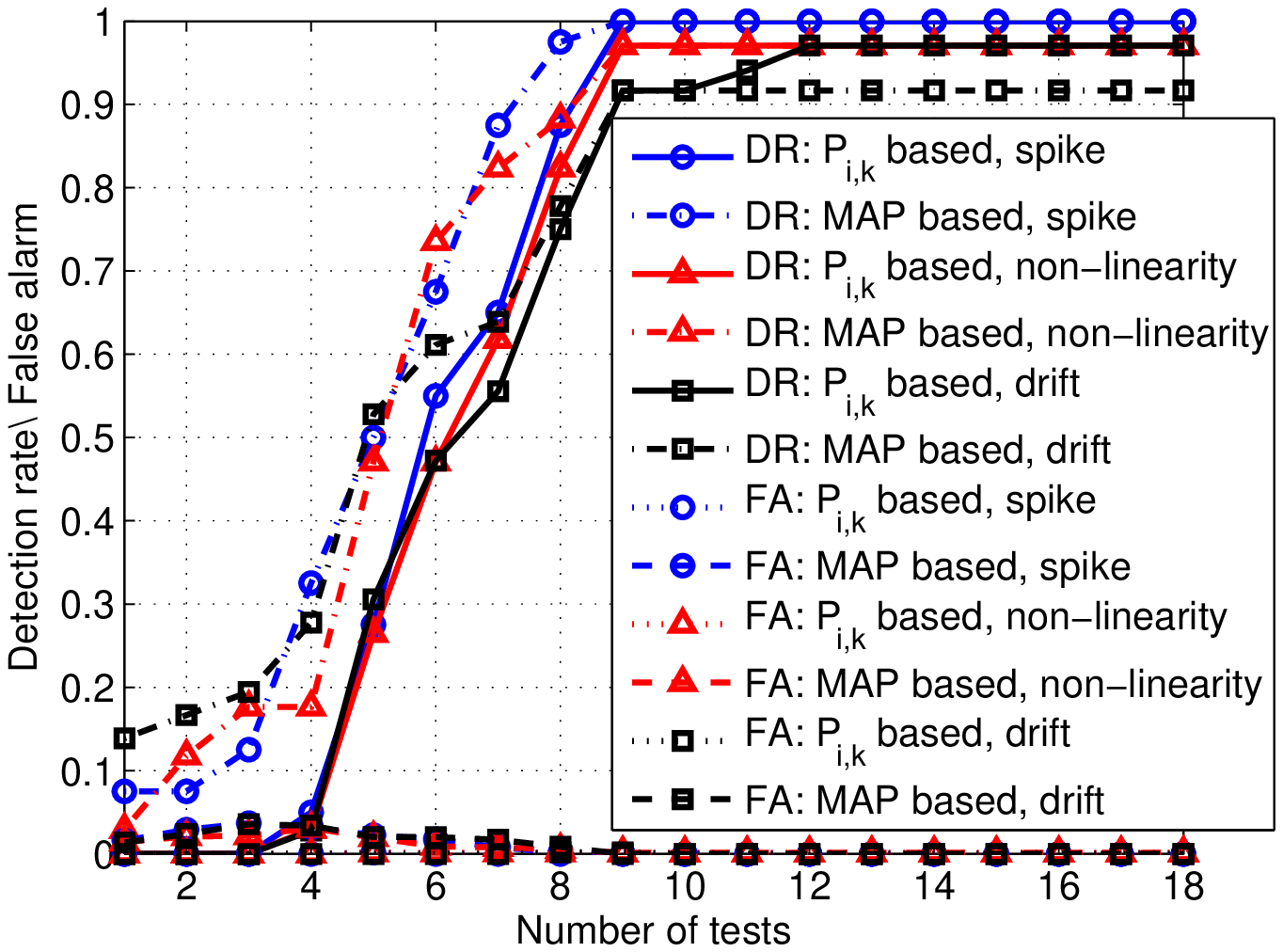}
\caption{The fault detection performance of MAP decoder and the $P_{i,k}$ based decoder.}
\label{fig:MAP_VS_P0}
\end{minipage}
\end{figure}

%

Fig. \ref{fig:BGT_VS_CGT} shows the performance comparison between BGT (with the $P_{i,k}$ decoder) and CGT. Fig. \ref{fig:MAP_VS_P0} compares the two decoders for  BGT: the MAP and the $P_{i,k}$ decoder. Both comparisons are evaluated on the same New Carquinz Bridge data, where 2 out of 18 sensors are faulty. For BGT the initial prior $P_{i,0}$ is set to 2/18 for each sensor. The first test pool is randomly generated with each sensor having probability 1/2 of being selected.  
The group test error \rev{$\alpha$ and $\beta$ are set to $0.01$}. When the $P_{i,k}$ decoder is used, sensor $i$ is regarded as faulty when the corresponding $P_{i,k}$ is smaller than $0.2$. The results are obtained from 50 random runs using the same setup as in the CGT evaluation in Section \ref{sec:CGT_result}. 

As shown in Fig. \ref{fig:BGT_VS_CGT}, BGT with the $P_{i,k}$ decoder outperforms  CGT on detecting all types of faults (when the number of tests $>6$).  BGT generally requires 3-4 fewer tests than the non-adaptive CGT for 80\% detection rate. Moreover, BGT uses 8 fewer tests to reach the saturation accuracy which is about 50\% improvement over CGT. The false alarm rates are similar ($<1\%$) for detecting different type of faults.
\rev{The improvement is primarily due to two sources:  BGT uses previous test results to design the next test, which leads to more effective tests; BGT is more conservative in deciding the sensor state (normal vs. faulty) and thus more robust when the group test is incorrect. }

For detecting the faulty sensor 11 shown in Fig. \ref{fig:faulty_sensor_example}, the initial prior $P_{i,0}$ is set to $1/18$ for all $i$. By selecting the first test pool $\Phi_1$ randomly, the BGT method is able to achieve $56\%$ detection rate ($0\%$ false alarm) when 5 tests are used and $100\%$ detection rate ($0\%$ false alarm) when 6 tests are used. The BGT algorithm saves 2 tests compared to the CGT algorithm for the same data set.

Fig. \ref{fig:MAP_VS_P0} compares the two state recover methods introduced in Section \ref{sec:methodology}. On average the MAP method is able to save one test for achieving the same accuracy as the $P_{i,k}$-based method. However, the MAP method has higher false alarm when the number of tests falls below 7. \rev{Also, the $P_{i,k}$-based method is preferred for large scale networks due to its low complexity.}
\rev{Note that neither decoding method requires the knowledge of $d$, the maximum number of faulty sensors. This is a significant benefit over CGT if $d$ is difficult to estimate.  CGT is not able to get correct result if $d$ is underestimated, due to the $d$-disjunct matrix requirement. Consequently, if $d$ is unknown then an overestimate is recommended for CGT, which then leads to an over-provisioning of the number of tests.}

%

\subsection{Performance of BGT in larger scale systems}
We next evaluate the performance of BGT in a large scale network (1000 sensors) and examine how it varies with the number of faulty sensors and group test error probabilities. 
A comparison between BGT and the divide-and-conquer adaptive group testing method proposed in \cite{hwang1972binary} is presented. \rev{We note that Hwang's method is designed for noiseless group test systems so it is not expected to work well with noisy group tests.  Nevertheless, it is meaningful to compare the two and quantify the difference under both noisy and noiseless conditions.} We also address the common prior initialization problem in Bayesian inference which also applies to BGT.  

For lack of real data on large networks, the experiments and results presented in this section are simulation based.  Out of the 1000 sensors, $d$ are randomly chosen and labeled as faulty. A group test result is first determined by whether the test pool contains any faulty sensors and then randomized according to the error model $\alpha=\beta$, i.e., with probability $\alpha$, the test result is flipped. In other words, we do not actually perform Kalman filtering based detection in this set of experiments, but its effect is simulated via this error model.

Hwang's method is based on the well-known binary search \cite{hwang1972binary}, whereby the network is first divided into 2 groups of equal size, and each is subject to the same group test process. If the result is negative, then all sensors in that group are declared normal removed from further testing; if the result is positive, then the group is further divided  into two smaller groups of equal size and the same process repeats until all faulty sensors have been identified.  Hwang's method has the following improvement compared to the standard  binary search.  It assumes knowledge on the number of faulty sensors $d$ (or an upper bound on $d$), and uses $d$ to determine the size of a group.  Specifically, when $d$ is small compared to the total number of uncertain sensors, a large test pool is used.  The idea is that upon a negative result a large number of sensors may be declared normal, and a new test pool can be selected from the remaining uncertain sensors; if the result is positive, the next test pool is generated randomly from the entire set of uncertain sensors, including the pool just tested positive.  Finally, when the number of remaining faulty sensors ($d$ minus the number of detected faulty sensors) is larger than half of the number of remaining uncertain sensors, the test is performed on an individual basis.

\rev{Clearly as mentioned, Hwang's method is designed for error-free group tests, so it does not handle errors well.  In particular, if a positive group is mistakenly detected as negative, this method will declare all faulty sensors in this group as normal and no further tests will be performed on them.  By contrast, BGT only decreases the probability of each tested sensor being normal, and they may be tested again in the future.  The comparison study here thus mainly  serves to quantify the improvement we can achieve when taking test errors into account.}

\begin{figure}[ht]
\centering
\begin{minipage}[b]{0.45\linewidth}
\includegraphics[width=7.5cm]{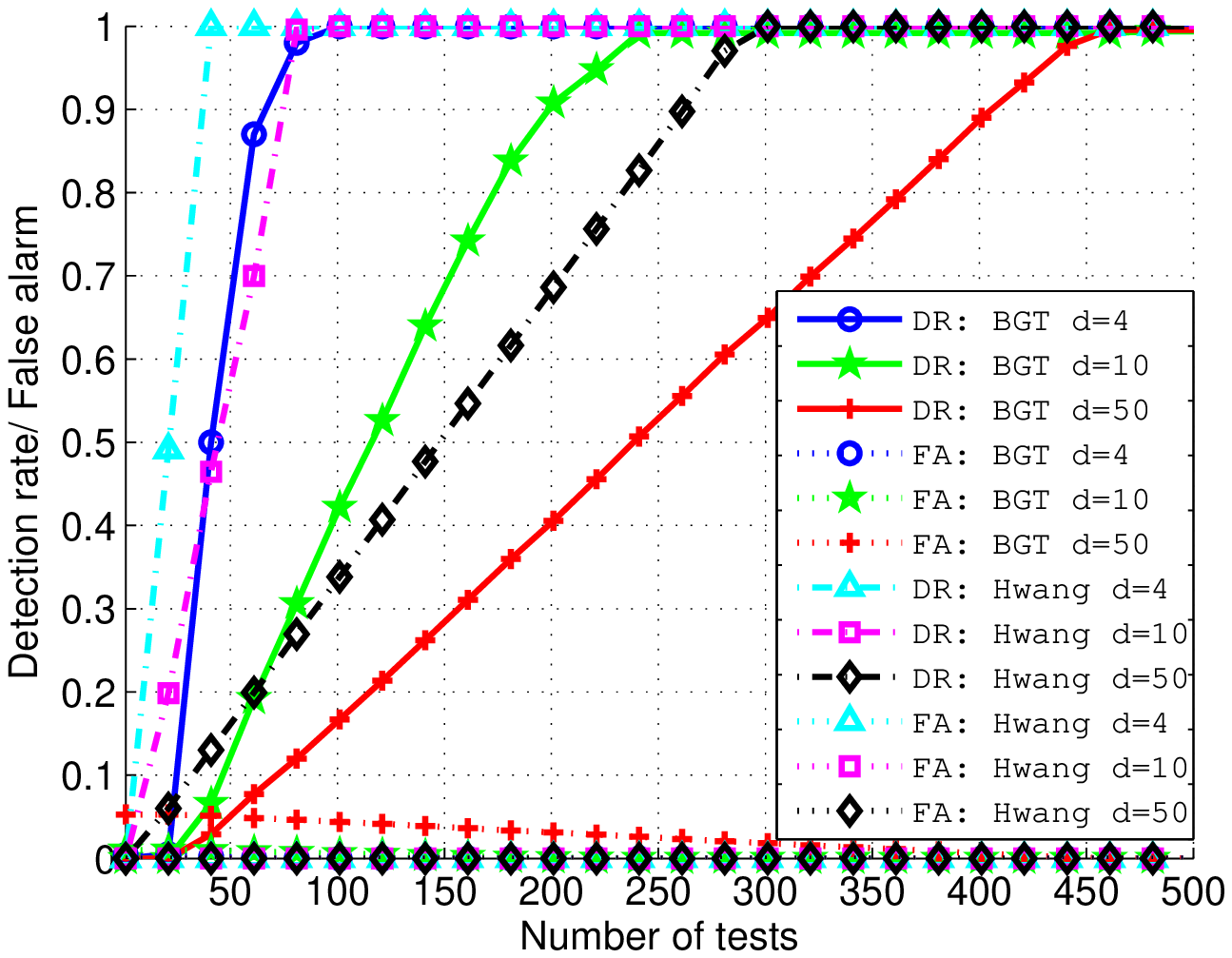}
\caption{The comparison of BGT and Hwang's methods with $\alpha=0$.}
\label{fig:BGT_VS_Hwang_0noise}
\end{minipage}
\quad
\begin{minipage}[b]{0.45\linewidth}
\includegraphics[width=7.5cm]{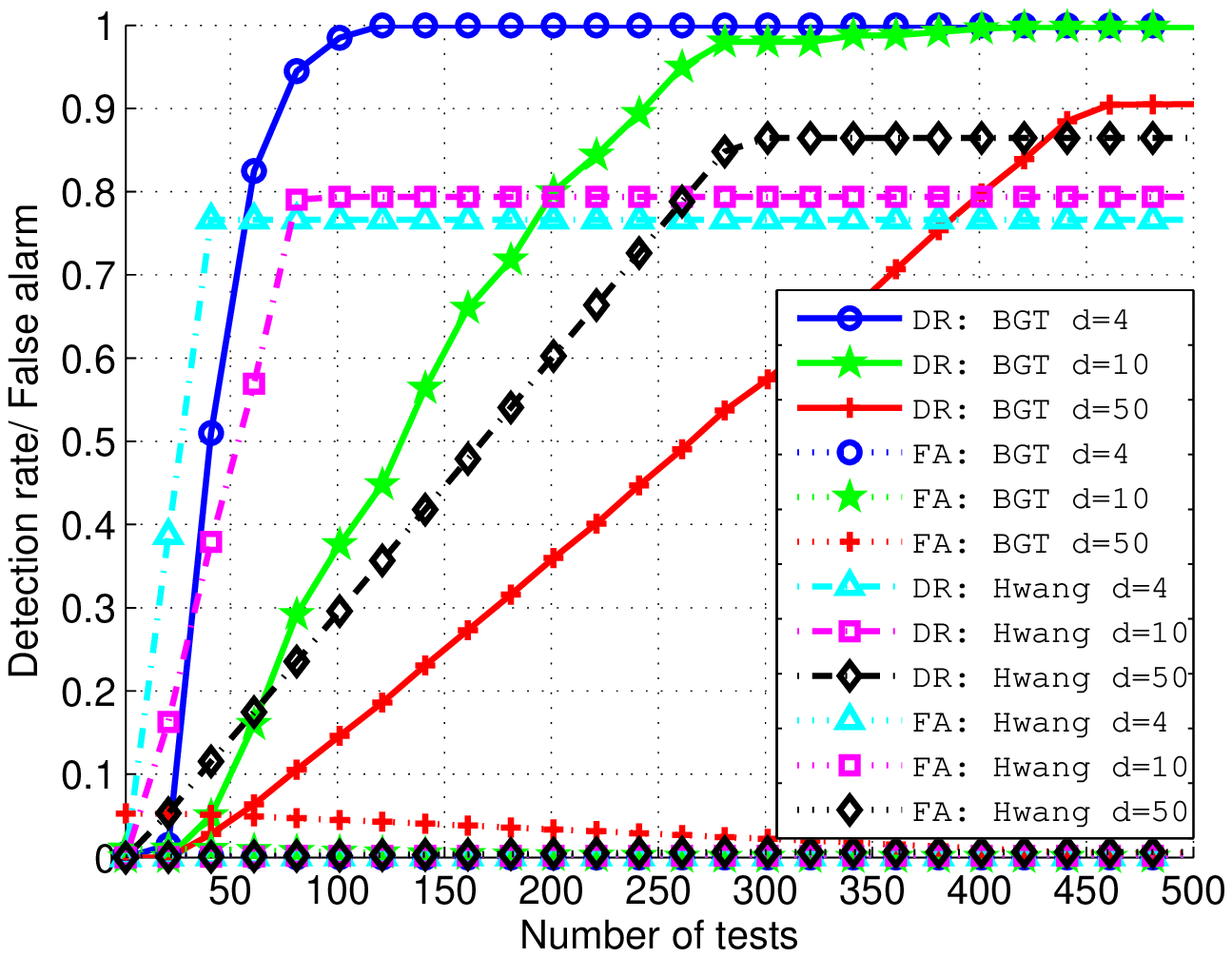}
\caption{The comparison of BGT and Hwang's methods with $\alpha=0.03$.}
\label{fig:BGT_VS_Hwang_003noise}
\end{minipage}
\end{figure}

\begin{figure}[ht]
\centering
\begin{minipage}[b]{0.45\linewidth}
\includegraphics[width=7.5cm]{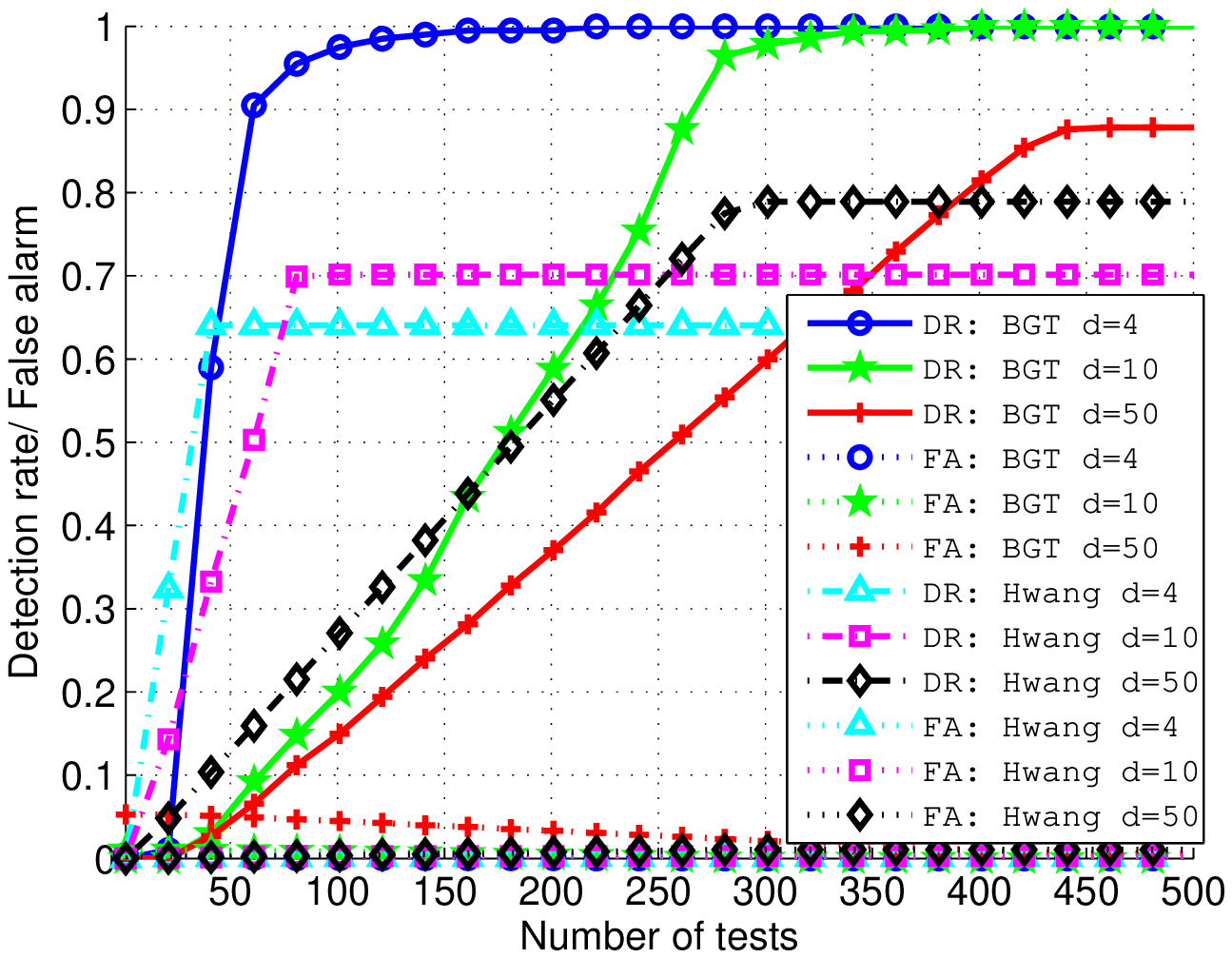}
\caption{The comparison of the BGT method and Hwang's methods with $\alpha=0.05$.}
\label{fig:BGT_VS_Hwang_005noise}
\end{minipage}
\quad
\begin{minipage}[b]{0.45\linewidth}
\includegraphics[width=7.5cm]{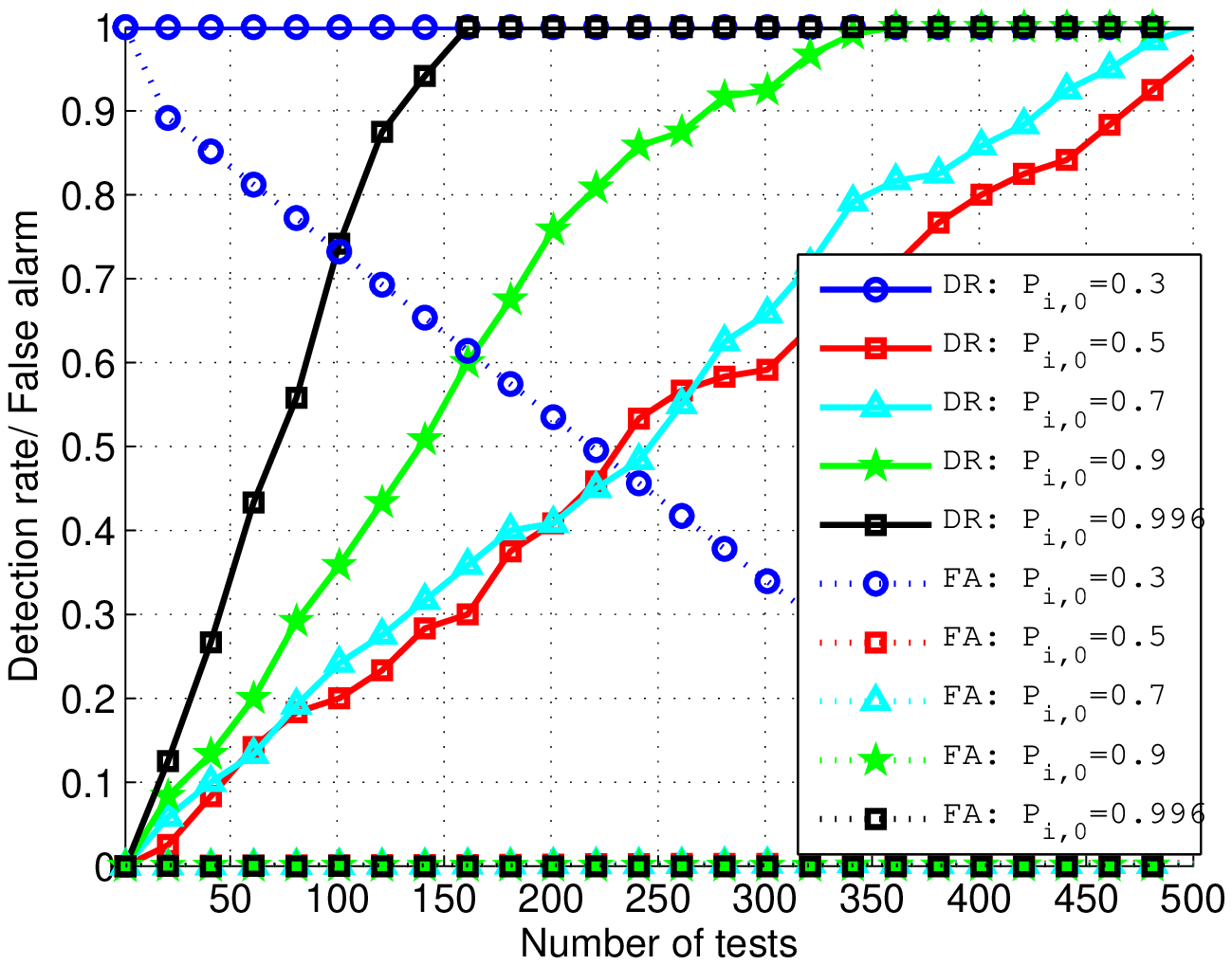}
\caption{The performance of BGT under different initial priors $P_{i,0}$ with the first test pool  selected randomly.}
\label{fig:prior_k_1}
\end{minipage}
\end{figure}

Figs. \ref{fig:BGT_VS_Hwang_0noise}-\ref{fig:BGT_VS_Hwang_005noise} show the performance of BGT and Hwang's method ($d=\{4, 10, 50\}$ respectively) under various group test error rates ($\alpha$).  When group tests are error-free (Fig. \ref{fig:BGT_VS_Hwang_0noise}), Hwang's method is able to achieve accurate results with fewer tests than BGT.  As expected, when group tests are noisy ($\alpha=0.03$ in Fig. \ref{fig:BGT_VS_Hwang_003noise} and $\alpha=0.05$ in Fig. \ref{fig:BGT_VS_Hwang_005noise}), BGT performs better while Hwang's method deteriorates rapidly.

%

\begin{figure}[ht]
\centering
\begin{minipage}[b]{0.45\linewidth}
\includegraphics[width=7.5cm]{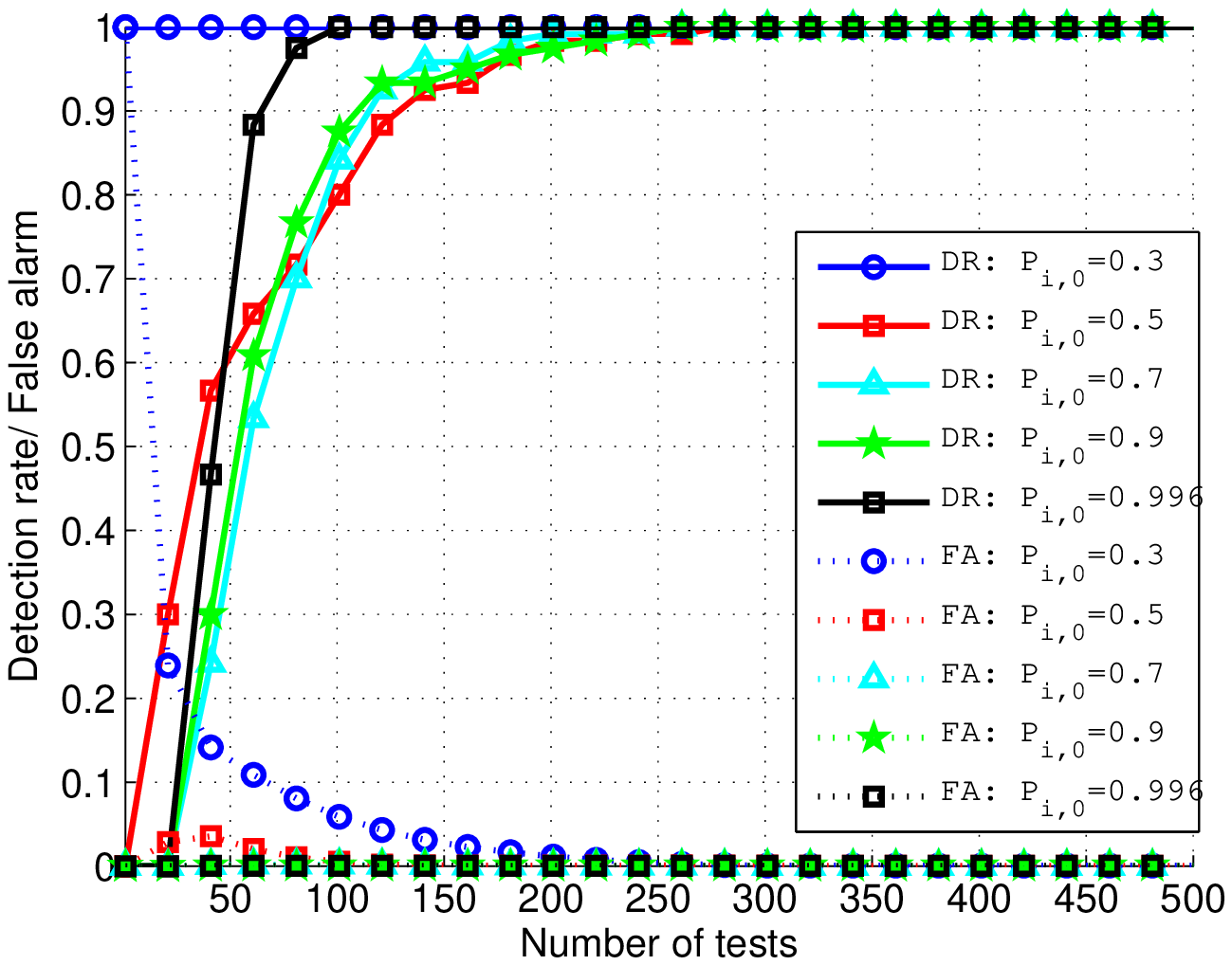}
\caption{The performance of BGT under different priors $P_{i,0}$ with the first 25 test pools selected randomly.}
\label{fig:prior_k_25}
\end{minipage}
\quad
\begin{minipage}[b]{0.45\linewidth}
\includegraphics[width=7.5cm]{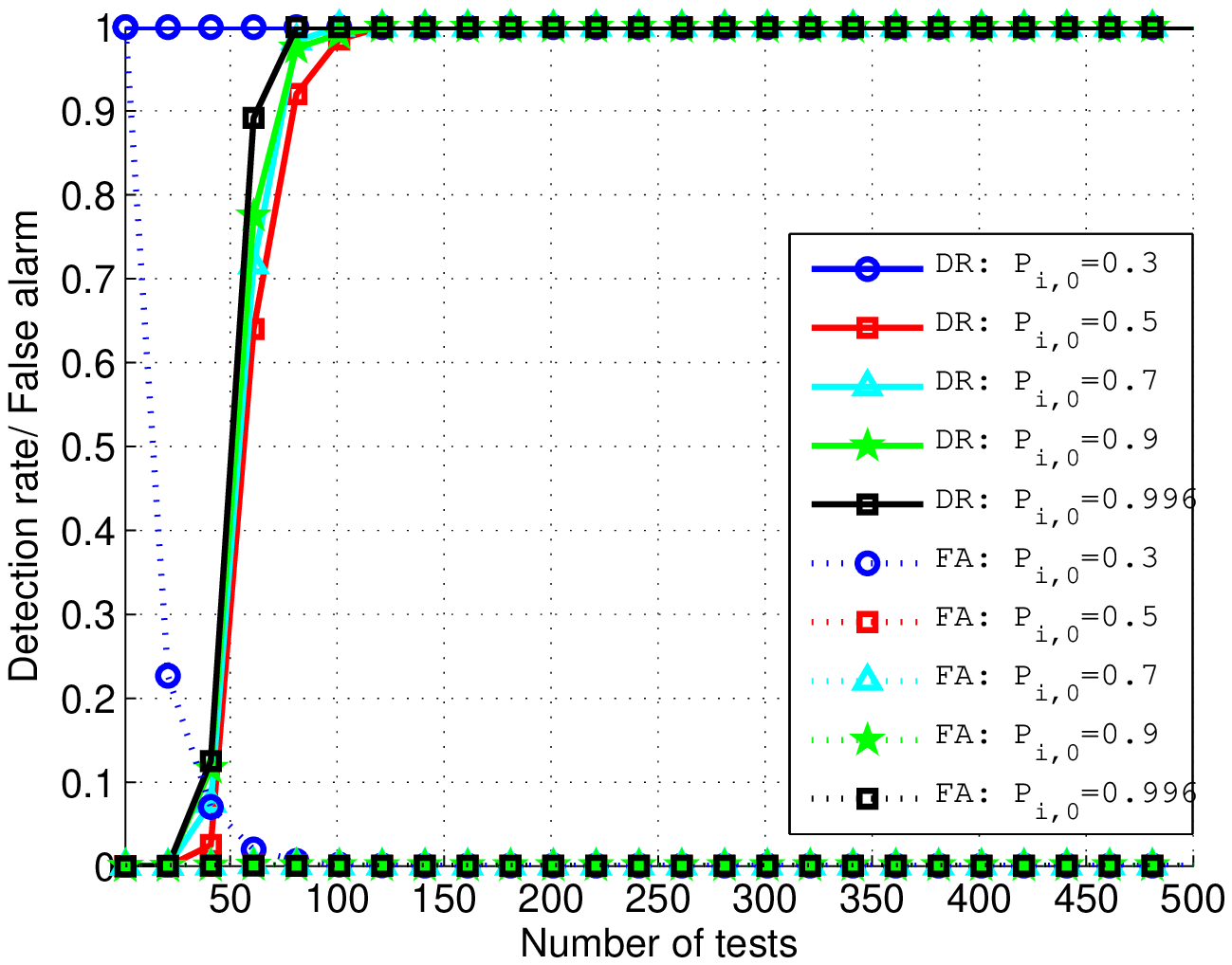}
\caption{The performance of BGT under different priors $P_{i,0}$ with the first 50 test pools selected randomly.}
\label{fig:prior_k_50}
\end{minipage}
\end{figure}

A common challenge to most Bayesian inference based methods is the selection of the prior on the hypothesis. 
Under BGT, the prior probability $P_{i,0}$ is required for designing a test pool. Fig. \ref{fig:prior_k_1} shows the result of using different priors ($P_{i,0}=\{0.3, 0.5, 0.7, 0.9, 0.96\}, \forall i$) when $d=4$ in a 1000-sensor network with $\alpha=0$. The case $P_{i,0}=0.96$ represents the correct prior. The figure shows that the performance is highly sensitive to the selection of the initial prior. However, this effect can be alleviated by choosing the first set of test pools randomly.  We see that when the first 25 test pools are randomly selected (each sensor has probability 1/2 to be selected), the difference in performance between different initial priors are significantly reduced (Fig. \ref{fig:prior_k_25}); when we increase this number to 50 tests (Fig. \ref{fig:prior_k_25}), this difference is largely eliminated.  Thus this random selection at the beginning serves as a very simple yet effective way to counter possible bad priors.  It may be seen as a form of {\em exploration} (random sampling) prior to {\em exploitation} (adaptive selection). 


%
%

To summarize, BGT is able to achieve the same performance as CGT with fewer tests, 
and is well suited for noisy group tests.  Furthermore, it does not require knowledge on$d$  when compared to CGT and Hwang's method. However, the adaptive design process prevents the use of parallel computing, which is viable for CGT. Therefore, CGT may actually have shorter run time if parallel computing is used.  

\section{The Design and Performance of KF-BGT}

Standard group tests are modeled as boolean operations.  While both CGT and BGT work with noisy group tests by modeling it as boolean operations with an error probability, they do not take into account other possible features of the group tests.  In our case, the group tests are given by the Kalman filtering based detection procedure, whose accuracy depends on not only the system model but also the test pools.  This suggests that a better understanding of the relationship between the detection procedure and the test pool design may allow us to further improve the design of the test pools and in turn the accuracy of the method.  This is the subject of investigation in this section.

\begin{figure}[ht]
\centering
\includegraphics[width=7.5cm]{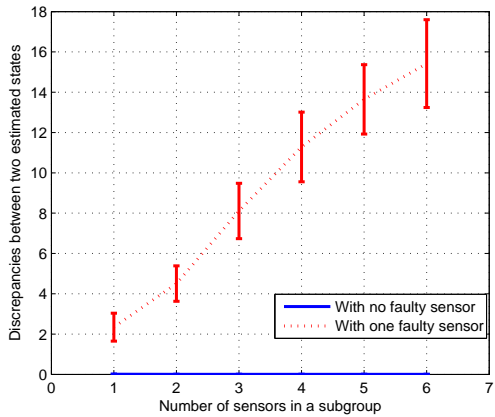}
\caption{The discrepancies in state estimates under different group sizes.}
\label{fig:EstimationGap_faulty_VS_nofault}
\end{figure}

We note that the Kalman filter estimates the state of a system based on the system model and the measurements from the sensors. As system identification method is used to obtain the system model, the model accuracy depends on the model order (the size of the system state, $S$) used. A higher order model generally gives better model accuracy (before over-fitting occurs) but it also requires more computational resources for the state estimation.  
The dependence of the state estimate accuracy on the size of the test group is shown in Fig. \ref{fig:EstimationGap_faulty_VS_nofault}.  
In this experiment, subgroups of different sizes are used to estimates the system state. For each group size, the discrepancy $|S_A-S_B|_\infty$ is recorded between having no faulty sensors in the subgroups and having one faulty sensor in one of the subgroups. When there are no faulty sensors, the discrepancy $|S_A-S_B|_\infty$ is very close to zero. \rev{On the other hand, $|S_A-S_B|_\infty$ is significantly larger with the presence of a single faulty sensor and increases with the group size.  This means that if a uniform detection threshold is used, then different group sizes will result in significantly different detection error (i.e., group test error) probabilities.  This further suggests that it would be desirable to maintain the same group sizes for the state estimate so as to keep the error probability constant and also to facilitate the choice of an optimal detection threshold.}
%

\begin{table}[h]
\caption{State estimate discrepancy $|S_A-S_B|_\infty$ under various faulty sensor distributions. (G: Number of good sensors, F: Number of faulty sensor)} 
\centering 
\begin{tabular}{c c c ||c c c c} 
\hline\hline 
\multicolumn{3}{c}{8 sensors} &\multicolumn{3}{c}{10 sensors}\\
\hline
\multicolumn{2}{c|}{Sensor distribution} &Discrepancy&\multicolumn{2}{c|}{Sensor distribution} &Discrepancy&\\
\hline
\multicolumn{1}{c|}{A:0G 4F}&\multicolumn{1}{c|}{B:4G 0F}&8.29&\multicolumn{1}{c|}{A:0G 6F}&\multicolumn{1}{c|}{B:6G 0F}&10.78&\\
\multicolumn{1}{c|}{A:1G 3F}&\multicolumn{1}{c|}{B:3G 1F}&23.88&\multicolumn{1}{c|}{A:1G 5F}&\multicolumn{1}{c|}{B:5G 1F}&26.73&\\
\multicolumn{1}{c|}{A:2G 2F}&\multicolumn{1}{c|}{B:2G 2F}&41.10&\multicolumn{1}{c|}{A:2G 4F}&\multicolumn{1}{c|}{B:4G 2F}&46.19&\\
\multicolumn{1}{c|}{}&\multicolumn{1}{c|}{}&&\multicolumn{1}{c|}{A:3G 3F}&\multicolumn{1}{c|}{B:3G 3F}&67.01&\\
\multicolumn{1}{c|}{A:4G 0F}&\multicolumn{1}{c|}{B:4G 0F}&7E-4&\multicolumn{1}{c|}{A:6G 0F}&\multicolumn{1}{c|}{B:6G 0F}&5E-4&\\
\hline
\end{tabular}
\label{tab:Faulty_sensor_dist}
\end{table}

We next examine the distribution of faulty sensors between two subgroups used in the filtering detection. Table \ref{tab:Faulty_sensor_dist} shows the state estimate discrepancy under various faulty sensor distribution in each subgroup. These results show that the discrepancy is highest when faulty sensors are evenly distributed between the two groups, e.g., having a faulty sensor in each subgroup is better than allocating two sensors in one subgroup as the larger discrepancy makes the detection more accurate.

Based on the above empirical observations, we propose the Kalman filtering (KF)-enhanced group test (KF-BGT) that uses the following rule in addition to the operation of BGT: after a new test pool has been selected using BGT, divide it \emph{evenly} into two subgroups.  If the there are fewer than 3 sensors in a subgroup, 
then sensors with high probability of being normal outside the test pool are added to the subgroups before performing Kalman filtering.

\begin{figure}[ht]
\centering
\includegraphics[width=7.5cm]{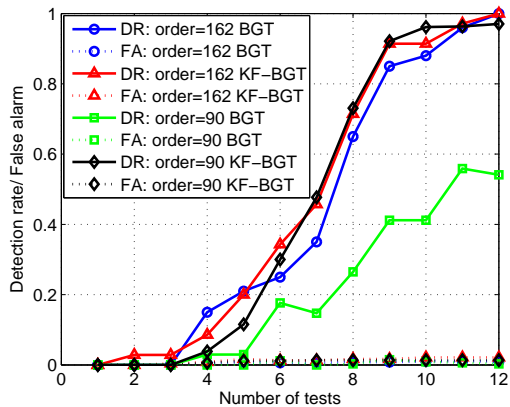}
\caption{The performance of BGT and KF-BGT under different model orders.}
\label{fig:BGT_VS_KFBGT}
\end{figure}

The differences in performance with and without the added sensor distribution step is illustrated in the following experiment. Fig. \ref{fig:BGT_VS_KFBGT} shows the detection rate and false alarm for non-linearity fault under BGT and KF-BGT. The performance is evaluated under both order 162 and order 90 system models. The performance of BGT  declines significantly under a less accurate system model (smaller model order). In contrast, the performance of KF-BGT only deteriorate slightly, thus it improves upon BGT significantly  when the system model is less accurate.  This shows that the sensor distribution makes the resulting method highly robust against the quality of the system model.

\section{Conclusion}
This study introduced group-testing-based sensor fault detection methods for sensor networks when faulty sensors are rare. We proposed a Kalman filter based group test method which is able to evaluate a random group of sensors and determine whether any faulty sensor is contained in the group. We presented a combinatorial group testing (CGT) method 
and a Bayesian group testing (BGT) method, which is adaptive and particularly suitable when group tests are noisy.   We also show how the computation involved in BGT can be made tractable for large networks.

Both CGT and BGT are evaluated by using a set of real vibration data collected by sensors deployed on the New Carquinez Bridge. Results show that both methods are able to reduce the number of required tests when faulty sensors are rare, compared to non-group testing methods. In addition, BGT uses fewer tests than CGT as it exploits previous test results in a sequential setting. 
We further propose an enhancement to BGT (KF-BGT) by taking into account features of the Kalman filtering process when determining subgroups in a single group test.


\bibliographystyle{ieeetr}
\bibliography{BGT_FD}

\begin{thebibliography}{10}

\bibitem{lo2013efficient}
C.~Lo, M.~Liu, J.~P. Lynch, and A.~C. Gilbert, ``Efficient sensor fault
  detection using combinatorial group testing,'' in {\em Distributed Computing
  in Sensor Systems (DCOSS), 2013 IEEE International Conference on},
  pp.~199--206, IEEE, 2013.

\bibitem{lynch2007overview}
J.~Lynch, ``An overview of wireless structural health monitoring for civil
  structures,'' {\em Philosophical Transactions of the Royal Society A:
  Mathematical, Physical and Engineering Sciences}, vol.~365, no.~1851,
  pp.~345--372, 2007.

\bibitem{WSN_agriculture}
T.~Wark, P.~Corke, P.~Sikka, L.~Klingbeil, Y.~Guo, C.~Crossman, P.~Valencia,
  D.~Swain, and G.~Bishop-Hurley, ``Transforming agriculture through pervasive
  wireless sensor networks,'' {\em Pervasive Computing, IEEE}, vol.~6, pp.~50
  --57, april-june 2007.

\bibitem{cartel}
B.~Hull, V.~Bychkovsky, Y.~Zhang, K.~Chen, M.~Goraczko, A.~Miu, E.~Shih,
  H.~Balakrishnan, and S.~Madden, ``Cartel: a distributed mobile sensor
  computing system,'' in {\em Proceedings of the 4th international conference
  on Embedded networked sensor systems}, pp.~125--138, ACM, 2006.

\bibitem{Ramanathan:2005:failure:sensor}
N.~Ramanathan, K.~Chang, R.~Kapur, L.~Girod, E.~Kohler, and D.~Estrin,
  ``Sympathy for the sensor network debugger,'' in {\em Proceedings of the 3rd
  international conference on Embedded networked sensor systems}, SenSys '05,
  (New York, NY, USA), pp.~255--267, ACM, 2005.

\bibitem{Staddon:2002:failednode:tracing}
J.~Staddon, D.~Balfanz, and G.~Durfee, ``Efficient tracing of failed nodes in
  sensor networks,'' in {\em Proceedings of the 1st ACM international workshop
  on Wireless sensor networks and applications}, WSNA '02, (New York, NY, USA),
  pp.~122--130, ACM, 2002.

\bibitem{Ruiz:2004:Event:Driven}
L.~B. Ruiz, I.~G. Siqueira, L.~B.~e. Oliveira, H.~C. Wong, J.~M.~S. Nogueira,
  and A.~A.~F. Loureiro, ``Fault management in event-driven wireless sensor
  networks,'' in {\em Proceedings of the 7th ACM international symposium on
  Modeling, analysis and simulation of wireless and mobile systems}, MSWiM '04,
  (New York, USA), pp.~149--156, ACM, 2004.

\bibitem{kobayashi2003application}
T.~Kobayashi and D.~Simon, ``Application of a bank of kalman filters for
  aircraft engine fault diagnostics,'' tech. rep., DTIC Document, 2003.

\bibitem{dalin:kalman}
R.~Da and C.~Lin, ``Sensor failure detection with a bank of kalman filters,''
  in {\em American Control Conference, 1995. Proceedings of the}, vol.~2,
  pp.~1122--1126, IEEE, 1995.

\bibitem{Li:iomodel}
Z.~Li, B.~Koh, and S.~Nagarajaiah, ``Detecting sensor failure via decoupled
  error function and inverse input--output model,'' {\em Journal of engineering
  mechanics}, vol.~133, no.~11, pp.~1222--1228, 2007.

\bibitem{Ricquebourg:TBM}
V.~Ricquebourg, D.~Menga, M.~Delafosse, B.~Marhic, L.~Delahoche, and
  A.~Jolly-Desodt, ``Sensor failure detection within the tbm framework: A
  markov chain approach,'' in {\em Proceedings of IPMU}, vol.~8, p.~323, 1991.

\bibitem{Lo:2011:reference_free_FD}
C.~Lo, J.~Lynch, and M.~Liu, ``Reference-free detection of spike faults in
  wireless sensor networks,'' in {\em Resilient Control Systems (ISRCS), 2011
  4th International Symposium on}, pp.~148 --153, aug. 2011.

\bibitem{Ding:Localized_F_Tolerant}
M.~Ding, D.~Chen, K.~Xing, and X.~Cheng, ``Localized fault-tolerant event
  boundary detection in sensor networks,'' in {\em INFOCOM 2005. 24th Annual
  Joint Conference of the IEEE Computer and Communications Societies.
  Proceedings IEEE}, vol.~2, pp.~902 -- 913 vol. 2, march 2005.

\bibitem{Chen:distributed}
J.~Chen, S.~Kher, and A.~Somani, ``Distributed fault detection of wireless
  sensor networks,'' in {\em Proceedings of the 2006 workshop on Dependability
  issues in wireless ad hoc networks and sensor networks}, pp.~65--72, ACM,
  2006.

\bibitem{koushanfar2003line}
F.~Koushanfar, M.~Potkonjak, and A.~Sangiovanni-Vincentelli, ``On-line fault
  detection of sensor measurements,'' in {\em Sensors, 2003. Proceedings of
  IEEE}, vol.~2, pp.~974--979, IEEE, 2003.

\bibitem{example_high_densityWSN}
S.~Cho and A.~Chandrakasan, ``Energy efficient protocols for low duty cycle
  wireless microsensor networks,'' in {\em Acoustics, Speech, and Signal
  Processing, 2001. Proceedings. (ICASSP '01). 2001 IEEE International
  Conference on}, vol.~4, pp.~2041 --2044 vol.4, 2001.

\bibitem{blough1989fault}
D.~Blough, G.~Sullivan, and G.~Masson, ``Fault diagnosis for sparsely
  interconnected multiprocessor systems,'' in {\em Fault-Tolerant Computing,
  1989. FTCS-19. Digest of Papers., Nineteenth International Symposium on},
  pp.~62--69, IEEE, 1989.

\bibitem{goodrich2006efficient}
M.~T. Goodrich and D.~S. Hirschberg, ``Efficient parallel algorithms for dead
  sensor diagnosis and multiple access channels,'' in {\em Proceedings of the
  eighteenth annual ACM symposium on Parallelism in algorithms and
  architectures}, pp.~118--127, ACM, 2006.

\bibitem{tosic2012distributedGT}
T.~Tosic and P.~Frossard, ``Distributed group testing detection in sensor
  networks,'' in {\em Acoustics, Speech and Signal Processing (ICASSP), 2012
  IEEE International Conference on}, pp.~3097--3100, IEEE, 2012.

\bibitem{hwang1972binary}
F.~Hwang, ``A method for detecting all defective members in a population by
  group testing,'' {\em Journal of the American Statistical Association},
  vol.~67, no.~339, pp.~605--608, 1972.

\bibitem{allemann2003BGT_split}
A.~Allemann, {\em Improved upper bounds for several variants of group testing}.
\newblock PhD thesis, Universit{\"a}tsbibliothek, 2003.

\bibitem{du1993combinatorial}
D.~Z. Du and F.~Hwang, {\em Combinatorial group testing and its applications}.
\newblock World Scientific, 1993.

\bibitem{malloy2012near_O_CS}
M.~L. Malloy and R.~D. Nowak, ``Near-optimal adaptive compressed sensing,'' in
  {\em Signals, Systems and Computers (ASILOMAR), 2012 Conference Record of the
  Forty Sixth Asilomar Conference on}, pp.~1935--1939, IEEE, 2012.

\bibitem{ji2008bayesianCS}
S.~Ji, Y.~Xue, and L.~Carin, ``Bayesian compressive sensing,'' {\em Signal
  Processing, IEEE Transactions on}, vol.~56, no.~6, pp.~2346--2356, 2008.

\bibitem{dorfman1943detection}
R.~Dorfman, ``The detection of defective members of large populations,'' {\em
  The Annals of Mathematical Statistics}, vol.~14, no.~4, pp.~436--440, 1943.

\bibitem{maybeck1979stochastic}
P.~Maybeck, {\em Stochastic Models, Estimation, and Control: Vol.: 1}.
\newblock Academic Press, 1979.

\bibitem{Gilbert_Recovering_simple_signals}
A.~Gilbert, B.~Hemenway, A.~Rudra, M.~Strauss, and M.~Wootters, ``Recovering
  simple signals,'' in {\em Information Theory and Applications Workshop (ITA),
  2012}, pp.~382 --391, feb. 2012.

\bibitem{Atia:2012:BooleanCS_meansurements}
G.~Atia and V.~Saligrama, ``Boolean compressed sensing and noisy group
  testing,'' {\em Information Theory, IEEE Transactions on}, vol.~58, pp.~1880
  --1901, march 2012.

\bibitem{cover2012elements}
T.~M. Cover and J.~A. Thomas, {\em Elements of information theory}.
\newblock John Wiley \& Sons, 2012.

\bibitem{katayama2005subspace}
T.~Katayama, {\em Subspace methods for system identification}.
\newblock Springer Verlag, 2005.

\bibitem{tong1998multichannel}
L.~Tong and S.~Perreau, ``Multichannel blind identification: From subspace to
  maximum likelihood methods,'' {\em Proceedings of the IEEE}, vol.~86, no.~10,
  pp.~1951--1968, 1998.

\bibitem{kurata2012NCQ}
M.~Kurata, J.~Kim, J.~Lynch, G.~v.~d. Linden, H.~Sedarat, E.~Thometz,
  P.~Hipley, and L.-H. Sheng, ``Internet-enabled wireless structural monitoring
  systems: Development and permanent deployment at the new carquinez suspension
  bridge,'' {\em Journal of Structural Engineering}, vol.~139, pp.~1688--1702,
  2012.

\end{thebibliography}

\end{document}